\documentstyle[amstex,amsfonts,aps,manuscript]{revtex}

\begin{document}
\draft
\author{Margaret Hawton\cite{email1} and William E. Baylis\cite{email2}}
\address{$^1$Department of Physics, Lakehead University,\\
Thunder Bay, Ontario, Canada P7B 5E1\\
$^2$Department of Physics, University of Windsor,\\
Windsor, Ontario, Canada }
\title{Photon position operators and localized bases}
\maketitle

\begin{abstract}
We extend a procedure for construction of the photon position operators with
transverse eigenvectors and commuting components [Phys. Rev. A {\bf 59}, 954
(1999)] to body rotations described by three Euler angles. The axial angle
can be made a function of the two polar angles, and different choices of the
functional dependence are analogous to different gauges of a magnetic field.
Symmetries broken by a choice of gauge are re-established by transformations
within the gauge group. The approach allows several previous proposals to be
related. Because of the coupling of the photon momentum and spin, our
position operator, like that proposed by Pryce, is a matrix that does not
commute with the spin operator. Unlike the Pryce operator, however, our
operator has commuting components, but the commutators of these components
with the total angular momentum require an extra term to rotate the matrices
for each vector component around the momentum direction. Several proofs of
the nonexistence of a photon position operator with commuting components are
based on overly restrictive premises that do not apply here.
\end{abstract}


\narrowtext

\section{Introduction}

Since the early days of quantum mechanics, it has been claimed that there is
no photon position operator with commuting components, and hence that a
basis of its localized eigenvectors, $\left| {\bf r}^{\prime }\right\rangle $%
, does not exist \cite{Pauli}. As a consequence, it is widely held that
there is no coordinate-space photon wave function, $\psi \left( {\bf r}%
^{\prime }\right) =\left\langle {\bf r}^{\prime }|\psi \right\rangle $ \cite%
{CohenTannoudji}. Wave functions for photons in momentum space are commonly
used, and it is there where the position operator would be defined. While it
may generally be possible to use a second-quantized formalism in which the
fields are operators in Fock space \cite{CohenTannoudji,Akh65}, the
usefulness of a coordinate-space photon wave function for describing such
phenomena as photon interference is sufficiently well recognized that a
number of authors have introduced versions of effective spatial wave
functions \cite{BialynickiBirula96,Sipe95}. The existence of a position
operator may also be important for the consistency of some work in the
second-quantized formalism.

A number of authors \cite{Flem65,Flem66,Heger74,Skag76,Ceg77,Janc77,Heger85}
have discussed and to some extent resolved well-known problems of Lorentz
covariance and causality that exist for any position operator in
relativistic quantum theory. However, there are additional problems for
massless particles of spin one or greater, and a number of ``proofs'' have
been presented of the nonexistence of a photon position operator in the
literature \cite{NewtonWigner,Wightman,JordanMukunda,Jadczyk73,Jordan,Mourad}%
. In spite of these, one of us has recently produced a counter example: a
Hermitian position operator with commuting components and localized
transverse eigenvectors \cite{Hawton1}. However, the asymmetric, singular
nature of the operator was puzzling, and questions concerning its
compatibility with work concluding the nonexistence of such operators were
not fully addressed.

In the present paper we attempt to resolve these issues. Our principal tool
is a generalization of the new position operator to include an arbitrary
axial rotation, which may be a function of the other two Euler angles that
parameterize the body rotation. This generalization provides insight into
the geometry underlying the operator and allows us to unify several previous
approaches.

While position operators can be defined in several ways, the one in Ref.\cite%
{Hawton1}, henceforth to be referred to as I, was constructed by requiring
its components to have eigenvectors transverse to the momentum and in the
directions of the momentum-space polar unit vectors. The position operator
thus obtained takes the form of a $3\times 3$ matrix and will be referred to
here as $\underline{{\bf r}}^{(0)},$ where the underscore denotes the matrix
character and{\rm \ }$(0)$ refers to use of the spherical polar basis
vectors $\bbox{\hat{\theta }}$ and $\bbox{\hat{\phi
}}$. The momentum-space operator \underline{${\bf r}$}$^{\left( 0\right) }$
can be expressed by a transformation that rotates the components of the
photon state function in momentum space to a fixed photon reference frame,
differentiates, and then rotates back to the lab frame. Since two angles
suffice to specify the direction of the photon momentum ${\bf p,}$ the
rotations require only two independent parameters.

Here we follow a more general approach by including, in addition to the
polar angles $\theta $ and $\phi $, the axial Euler angle, $\chi $.{\em \ }%
Let the unit vectors along the Cartesian axes be ${\bf e}_{1}$, ${\bf e}_{2}$
and ${\bf e}_{3}$. A rotation about ${\bf p}$ by an angle $\chi $ permits an
arbitrary choice of the transverse unit vectors ${\bf e}_{{\bf p}1}$ and $%
{\bf e}_{{\bf p}2}\,,$ which are obtained by a rotation from the fixed unit
vectors ${\bf e}_{1}$ and ${\bf e}_{2}$ in the given reference frame. The
same rotation is designed to take ${\bf e}_{3}$ to the momentum direction $%
{\bf \hat{p}=p/}p$, but ${\bf \hat{p}}$ depends only on the polar angles $%
\theta $ and $\phi ;$ it is independent of $\chi .${\bf \ }We show below
that if $\chi $ is independent of $\theta $ and $\phi ,$ the position
operator is just $\underline{{\bf r}}^{(0)}$ as found previously\cite%
{Hawton1}. However, we are also free to choose $\chi $ to be a function $%
\chi =\chi _{{\bf p}}\left( \theta ,\phi \right) $ of the polar angles. The
momentum-space position operator that corresponds to such a choice is 
\begin{equation}
\underline{{\bf r}}=\underline{{\bf r}}^{(0)}+\nabla \chi _{{\bf p}}
\label{rp}
\end{equation}%
in units with $\hbar =1$. (It is understood that the second term on the rhs
is multiplied by a unit $3\times 3$ matrix. To simplify notation, we do not
indicate unit matrices explicitly.) A change in the momentum-space gradient $%
\nabla \chi _{{\bf p}}(\theta ,\phi )$ is analogous to a gauge
transformation of \underline{${\bf r}$}. The singularity previously found in 
\underline{${\bf r}$}$^{\left( 0\right) }$ is analogous to that of a
monopole string and commonly arises in electromagnetic-like gauge
potentials. For any choice of $\chi _{{\bf p}},$ \underline{${\bf r}$} has
less symmetry than one would expect and under parity inversion it does not
simply change its sign. However, as we show below, the symmetry is restored
when one includes the full group of possible gauge transformations rather
than only a single choice of gauge.

In the case of massive particles of spin $S$, any component of the spin can
be used to define an eigenstate basis. The basis functions corresponding to
a space-fixed quantization direction span a rotationally invariant linear
manifold and carry irreducible representations of the Poincar\'{e} group.
Because the momentum-space basis states for massive particles have
directions in space that are invariant under infinitesimal translations, the
position operator can be identified with the Hermitian generator $i\nabla $
of such translations, where $\nabla $ is the momentum-space gradient. Since $%
i\nabla $ is a differential operator proportional to the unit matrix, it
commutes with spin matrices that represent space-fixed components of the
spin. For massless particles, on the other hand, representations of the
Poincar\'{e} group are reduced to the two irreducible representations
carried by helicity states in which the spin component along the momentum $%
{\bf p}$ is $\pm S\,$\cite{Wigner}. Since the two helicity subspaces are
invariant under the actions of the Poincar\'{e} group, the allowed spin
states for a massless particle with $S\geq 1$ form rotationally invariant
manifolds with fewer than the $2S+1$ independent elements required for
states quantized along a given spatial direction. As elaborated below (see
Section VI), this is important in understanding the relevance of the
nonexistence result of Newton and Wigner \cite{NewtonWigner}. The basis
vectors corresponding to the two allowed helicities are transverse to ${\bf 
\hat{p}}$ and therefore require a modified position operator, one that is no
longer independent of the spin.

The generator of rotations is, as usual, 
\begin{equation}
\underline{{\bf J}}=-i{\bf p\times }\nabla +\underline{{\bf S}},
\label{Rgenerator}
\end{equation}%
where the vector components of $\underline{{\bf S}}$ are the spin-one
matrices \underline{$S_{j}$} whose explicit elements depend of the basis
used for the state vectors (see discussion in the following section). The
term $-i{\bf p\times \nabla =L}$ is required in order to generate a rotation
of the argument of the wave function, whereas $\underline{{\bf S}}$
generates the rotation of its components. Our use of the sum (\ref%
{Rgenerator}) for photons is consistent with Bargmann and Wigner's \cite%
{BargmannWigner} proof that $\underline{{\bf J}}$ is the generator of
rotations for a particle of arbitrary mass.

For the position operator ${\bf r}$ to transform as a vector, it is
traditionally required that 
\begin{equation}
\left[ \underline{J_{j}},r_{k}\right] =i\varepsilon _{jkl}r_{l}\,,
\label{commJ}
\end{equation}%
where repeated indices are summed over. For massive particles, this
commutator follows directly from the assumptions that components of the
operator ${\bf r}$ commute with each other and with the spin matrices, and
that they satisfy the canonical commutator relations with the momentum,%
\begin{equation}
\left[ p_{j},r_{k}\right] =-i\delta _{jk}\,,
\end{equation}%
since then the components of either ${\bf L}$ or \underline{${\bf J}$}
rotate one component of ${\bf r}$ into another:%
\begin{eqnarray}
\left[ \underline{J_{j}},r_{k}\right] &=&\left[ L_{j},r_{k}\right]
=\varepsilon _{jmn}r_{m}\left[ p_{n},r_{k}\right] \\
&=&i\varepsilon _{jkm}r_{m}\,.
\end{eqnarray}%
Whereas this clearly applies to the position operator $i\nabla $ usually
chosen for massive particles, the photon position operators proposed by
Pryce \cite{Pryce} and one of us \cite{Hawton1} are matrices that do not
commute with the spin. As we discuss in more detail in the following
section, components of the Pryce position operator satisfy the commutation
relations (\ref{commJ}) but do not commute with each other. Consequently,
their eigenstates cannot have eigenvalues for more than one component of the
position operator and thus cannot fully define the position of a photon.

On the other hand, our position operator has commuting components, but at
the cost of the commutation relation (\ref{commJ}). For a given functional
form of $\chi _{{\bf p}},$ an infinitesimal rotation of \underline{${\bf r}$}
results in an incremental axial rotation of the matrix \underline{$r_{j}$}
for each component by $d\chi -d\chi _{{\bf p}}\,$, and this yields the extra
term derived below [see Eq.\thinspace (\ref{rJcomm})]. Recent proofs of the
nonexistence of a photon position operator and localized states have assumed
components $r_{j}$ that satisfy Eq.(\ref{commJ}) \cite{Jordan,Mourad}, thus
excluding the position operators discussed here.

As in I, the position operator is constructed with the requirement that its
components have transverse eigenvectors. It is related to the position
operator for massive particles, whose components have eigenvectors with
fixed directions in momentum space, by a transformation with the form of a
spin rotation and a dilation. As discussed more fully in the following
section, this similarity-like transformation includes a spin rotation $%
D\left( \phi ,\theta ,\chi \right) $ from fixed directions in the photon
reference frame to the longitudinal and transverse directions in the lab.
Because the longitudinal direction is independent of the axial angle $\chi ,$
a family of transformations with different functional dependencies $\chi
=\chi _{{\bf p}}\left( \theta ,\phi \right) $ are possible. We find that
this family of transformations includes ones that relate the Shirokov and
Lomont and Moses irreducible helicity representations to the Foldy form of
the Poincar\'{e} algebra. The use of Euler angles thus gives a unified
approach to the relationships among these representations and to the
position operators with transverse eigenvectors obtained in the present
paper.

In Section II we briefly review the Foldy and helicity representations of
the Poincar\'{e} operators. We then describe the relationship of the angular
momentum and boost operators to the Pryce \cite{Pryce} position operator. In
Section III the results of paper I are reviewed and then generalized to an
arbitrary functional dependence $\chi =\chi _{{\bf p}}\left( \theta ,\phi
\right) $ of the axial Euler angle on the polar angles of ${\bf \hat{p}}$.
The Poincar\'{e} group, the relationship of the position operator to the
angular momentum and to the boost operators, and the localized states are
then examined in Section IV. The connection of our work to Berry's phase is
treated in Section V, and the relation of string singularities to
nonintegrable angles is stressed. In Section VI, the consistency of our
position operator with nonexistence proofs is addressed, and its
implications for the photon wave function are briefly discussed. Details and
alternative derivations of some of the geometric arguments are presented in
the Appendices.

\section{Foldy and helicity representations of the Poincar\'{e} operators}

Foldy \cite{Foldy} wrote down the form of the ten generators of
inhomogeneous Lorentz transformations (the Poincar\'{e} operators) for
particles with spin $S$ and mass $m$ in the standard space-fixed
representation. His operators also exist in the $m\rightarrow 0$ limit, the
only case considered here, and in momentum space they comprise the momentum $%
{\bf p}$, the Hamiltonian $H=pc$, the total angular momentum operator ${\bf J%
}$ [Eq.(\ref{Rgenerator})], and the boost operator 
\begin{equation}
\underline{{\bf K}}=ip\nabla +{\bf \hat{p}}\times \underline{{\bf S}}\text{.}
\label{K}
\end{equation}%
(Foldy's symmetrized form of the $ip\nabla $ term was criticized by
Chakrabarti \cite{Chakrabarti}. The correct form is related to the
momentum-space normalization weight used for scalar products, which is
discussed toward the end of this section. Note that the boost generator $%
{\bf K}$ is ${\bf -N}$ in Chakrabarti's paper \cite{Chakrabarti}.)

The standard helicity representation was introduced by Lomont and Moses \cite%
{LomontMoses} to provide a realization of the generators for the zero-mass
case. The representation is distinct from the zero-mass limit of the Foldy
representation, given above, because the carrier space has been split into
invariant subspaces labeled by the two helicity components $\kappa =\pm S$.
Vector components of the spin ${\bf S}$ are referred to the photon reference
frame in which the momentum direction ${\bf \hat{p}}$ is ${\bf e}_{3}.$
Chakrabarti \cite{Chakrabarti} showed that the zero-mass limit of the Foldy
representation is related to the Lomont and Moses helicity representation by
the unitary transformation 
\begin{equation}
\underline{O}_{{\rm F}}=\underline{U}\,\underline{O}_{{\rm LM}}\underline{U}%
^{-1},  \label{FLM}
\end{equation}%
where \underline{$O$}$_{{\rm F}}$ is an operator in the Foldy
representation, \underline{$O$}$_{{\rm LM}}$ is the corresponding operator
in the helicity representation of Lomont and Moses, and the unitary
transformation has a matrix representation \underline{$U$} that we recognize
as the spin rotation%
\begin{equation}
\underline{U}=\exp (-i\theta \bbox{\hat{\phi }}\cdot \underline{{\bf S}}).
\label{Uh}
\end{equation}%
The transformation (\ref{FLM}) rotates $S_{3}$ directly into ${\bf S\cdot 
\hat{p}.}$ As above, $\theta $ and $\phi $ are the usual spherical polar
angles of ${\bf p}$ in momentum space and $\bbox{
\hat{\phi}}$ is the unit tangent vector in the direction of increasing $\phi 
$. It is perpendicular to both ${\bf e}_{3}$ and ${\bf \hat{p}\,.}$ In the
transformation (\ref{FLM}) and its inverse, it is important to recognize
that the unitary operator Eq.(\ref{Uh}) does not generally commute with the
momentum-space gradient $\nabla $ because of the dependence of $\theta $ and 
$\bbox{\hat{\phi}}$ on ${\bf \hat{p}.}$ The Hermitian generators of the
infinitesimal transformations of the Lomont and Moses helicity
representation are thus given by the Foldy representation and the inverse of
(\ref{FLM}) to be \cite{Chakrabarti} ${\bf p\equiv }\sum_{j}p_{j}{\bf e}%
_{j}, $ $H=pc$, 
\begin{equation}
\underline{{\bf J}}_{{\rm LM}}=-i{\bf p}\times \nabla {\bf +}\left( \frac{%
{\bf \hat{p}+e}_{3}}{1+\cos \theta }\right) \underline{S_{3}}\text{,}
\label{Jh}
\end{equation}%
and 
\begin{equation}
\underline{{\bf K}}_{{\rm LM}}=ip\nabla {\bf +}\left( \frac{{\bf \hat{p}%
\times e}_{3}}{1+\cos \theta }\right) \underline{S_{3}}  \label{Kh}
\end{equation}%
where the matrix \underline{$S_{3}$} can be replaced by $\kappa $ when
restricted to the invariant subspace of definite helicity.

A different helicity representation was obtained earlier by Shirokov \cite%
{Shirokov}. After conversion to polar coordinates, his angular-momentum and
boost operators are 
\begin{equation}
\underline{{\bf J}}_{{\rm Sh}}=-i{\bf p}\times \nabla {\bf +}\left( %
\bbox{\hat{\theta }}\cot \theta +{\bf \hat{p}}\right) \underline{S_{3}}
\label{JS}
\end{equation}%
and 
\begin{equation}
\underline{{\bf K}}_{{\rm Sh}}=ip\nabla +\bbox{\hat{\phi }}\cot \theta \,%
\underline{S_{3}}\text{.}  \label{KS}
\end{equation}%
The Bia\l ynicki-Birulas \cite{BB2,BialynickiBirula} derived results
equivalent to Eqs.(\ref{JS}) and (\ref{KS}) by considering the Maxwellian
momentum and angular-momentum tensors. They also found \cite%
{BialynickiBirula} a unitary matrix \underline{$W_{1}$} that relates
operators in the Foldy and Shirokov representations:%
\begin{equation}
\underline{O}_{{\rm F}}=\underline{W}_{1}\underline{O}_{{\rm Sh}}^{\left(
1\right) }\underline{W}_{1}^{-1}  \label{FShprime}
\end{equation}%
We can express their \underline{$W_{1}$} as the product of a transformation 
\underline{$T$} from a Cartesian to an angular-momentum basis and another
rotation 
\begin{equation}
\underline{D}\left( \phi ,\theta ,0\right) =\exp \left( -i\underline{S_{3}}%
\phi \right) \exp \left( -i\underline{S_{2}}\theta \right)
\end{equation}%
that takes ${\bf e}_{3}$ into ${\bf \hat{p}}$ : \underline{$W_{1}$}$=%
\underline{D}\left( \phi ,\theta ,0\right) \underline{T}=\underline{T}\,%
\underline{D}^{\left( 1\right) }\left( \phi ,\theta ,0\right) .$ The unitary
transformation 
\begin{equation}
\underline{T}=\frac{1}{\sqrt{2}}\left( 
\begin{array}{ccc}
1 & 0 & 1 \\ 
i & 0 & -i \\ 
0 & \sqrt{2} & 0%
\end{array}%
\right)  \label{T}
\end{equation}%
is required only because different bases for the momentum-space vectors have
been assumed in the two representations. Specifically in the transformation (%
\ref{FShprime}), the Foldy operator \underline{$O$}$_{{\rm F}}$ uses a
Cartesian basis, in which the unit vectors ${\bf e}_{1},{\bf e}_{2},{\bf e}%
_{3}$ are represented by column vectors with elements $\left( {\bf e}%
_{j}\right) _{k}=\delta _{jk}\,,$ whereas \underline{$O$}$_{{\rm Sh}%
}^{\left( 1\right) }$ uses an angular-momentum basis in which the complex
unit vectors ${\bf e}_{\kappa },\;\kappa =\pm 1,0,$ are column matrices with
elements $\left( {\bf e}_{\kappa }\right) _{k}=\delta _{\kappa ,k-2}\,.$
They are given in the Cartesian basis by the columns of \underline{$T$} (\ref%
{T}). The spin matrices are different in the two bases. In the Cartesian
basis, the matrices \underline{$S_{j}$} have elements $\left( S_{j}\right)
_{kl}=-i\varepsilon _{jkl}$ whereas in the spin-1 angular-momentum basis the
corresponding matrices are $\left( S_{3}^{\left( 1\right) }\right)
_{kl}=\left( 2-k\right) \delta _{kl},$ and $\left( S_{1}^{\left( 1\right)
}\pm iS_{2}^{\left( 1\right) }\right) _{kl}=\sqrt{2}\delta _{k,l\pm 1}\,.$
The body rotation matrices, expressed in terms of the Euler angles $\phi
,\theta ,\chi $ by%
\begin{equation}
\underline{D}(\phi ,\theta ,\chi )=\exp \left( -i\underline{S_{3}}\phi
\right) \exp \left( -i\underline{S_{2}}\theta \right) \exp \left( -i%
\underline{S_{3}}\chi \right) ,  \label{Drot}
\end{equation}%
have correspondingly different elements in the two bases. Explicit values in
the Cartesian basis are given in the following section and in Appendix A.
Elements of \underline{$D$}$^{\left( 1\right) }=\underline{T}^{-1}\underline{%
D}\,$\underline{$T$} in the angular-momentum basis have the standard form $%
D_{\kappa \kappa ^{\prime }}^{\left( 1\right) }=\exp \left( -i\kappa \phi
-i\kappa ^{\prime }\chi \right) d_{\kappa \kappa ^{\prime }}^{\left(
1\right) }\left( \theta \right) $, where the real matrix \underline{$d$}$%
^{\left( 1\right) }$ is \cite{BrinkSatchler}%
\begin{equation}
\underline{d}^{\left( 1\right) }\left( \theta \right) =\frac{1}{2}\left( 
\begin{array}{ccc}
\cos \theta +1 & \sqrt{2}\sin \theta & \cos \theta -1 \\ 
-\sqrt{2}\sin \theta & 2\cos \theta & -\sqrt{2}\sin \theta \\ 
\cos \theta -1 & \sqrt{2}\sin \theta & \cos \theta +1%
\end{array}%
\right) \,.
\end{equation}%
We will generally assume that the different representations both use the
Cartesian basis or both use an angular-momentum basis for their
momentum-space vectors, as is actually implicit in relation (\ref{FLM})
between the Foldy and Lomont-Moses representations. We can then omit the
factor \underline{$T$} and replace the transformation (\ref{FShprime}) by%
\begin{equation}
\underline{O}_{{\rm F}}=\underline{D}\left( \phi ,\theta ,0\right) 
\underline{O}_{{\rm Sh}}\underline{D}^{-1}\left( \phi ,\theta ,0\right) \,.
\label{FSh}
\end{equation}

The helicity representations of Lomont and Moses and of Shirokov are both
seen to be unitarily equivalent to the standard Foldy representation, and
the unitary transformations for both are rotations that rotate \underline{$%
S_{3}$} into \underline{${\bf S}$}${\bf \cdot \hat{p}.}$ It seems surprising
that the two helicity representations appear so different. As Chakrabarti %
\cite{Chakrabarti} points out, the Shirokov form of the operators ${\bf J}$
and ${\bf K}$ are of a singular nature that makes them appear unsatisfactory
for many uses. However, the Lomont and Moses form (\ref{Jh},\ref{Kh}) also
has first-order singularities at $\theta =\pi ,$ they are just better
disguised. The differences in the unitary transformations (\ref{FLM}) and (%
\ref{FSh}) is clearer if \underline{$U$} is expressed in terms of Euler
angles. Since 
\begin{equation}
\exp \left( -i\underline{S_{3}}\phi \right) \underline{S_{2}}\exp \left( i%
\underline{S_{3}}\phi \right) =\underline{S_{2}}\cos \phi -\underline{S_{1}}%
\sin \phi =\underline{{\bf S}}\cdot \hat{\phi},
\end{equation}%
then from the Euler-angle form (\ref{Drot}) we can equate%
\begin{equation}
\underline{U}=\underline{D}\left( \phi ,\theta ,-\phi \right) \,,
\end{equation}%
that is, the unitary transformation \underline{$U$} that rotates \underline{$%
S_{3}$} directly to \underline{${\bf S}$}${\bf \cdot \hat{p}}$ is just the
Euler-angle rotation \underline{$D$}$\left( \phi ,\theta ,0\right) $
required for the Shirokov form preceded by an axial rotation through the
angle $-\phi .$ It is important to note that the transformations involved
are {\em spin }rotations and therefore transform only spin components and
the vector components of the photon states. We explore a generalization of
such rotations further in the next section.

While the position operators are not explicitly required to complete the
Poincar\'{e} algebra, they are implicit in the angular momentum and boost
operators since, apart from internal degrees of freedom, it is the
components of ${\bf r}\times {\bf p}$ and $\frac{1}{2}(p{\bf r+r}p)$ that
generate rotations and boosts, respectively. One set of commonly accepted
photon position operators consists of components of the Pryce operator \cite%
{Pryce,Mourad}, 
\begin{equation}
\underline{{\bf r}}_{{\rm P}}=ip^{\alpha }\nabla p^{-\alpha }+\frac{{\bf %
p\times }\underline{{\bf S}}}{p^{2}}.  \label{rP}
\end{equation}%
The parameter $\alpha $ in the expression%
\begin{equation}
p^{\alpha }\nabla p^{-\alpha }=\nabla +\frac{\alpha }{p}\left[ p,\nabla %
\right] =\nabla -\frac{\alpha {\bf p}}{p^{2}}
\end{equation}%
depends on the integration weight used in the definition of the scalar
product. Pryce used the parameter value $\alpha =1/2$, which is appropriate
when the Hilbert-space states are electromagnetic fields, and this value was
implicitly assumed in Eqs.(\ref{K}), (\ref{Kh}), and (\ref{KS}).{\em \ }%
However, one needs $\alpha =-\frac{1}{2}$ when the vector potential \cite%
{Hawton2} is used for the photon states. Alternatively, a definition of
scalar product requiring $\alpha =0$ can be used \cite{Mourad,Akh65}. In the
following, we leave $\alpha $ unspecified.

The Pryce position operator \cite{Pryce} \underline{${\bf r}$}$_{{\rm P}}$
is based on expressions in terms of a classical energy-momentum tensor of a
noncovariant definition of the center of mass given by Fokker \cite{Fokker29}%
. The components of \underline{${\bf r}$}$_{{\rm P}}$ are $3\times 3$
matrices that do not commute with \underline{$S_{j}$}. When expressed in
terms of \underline{${\bf r}$}$_{{\rm P}},$ the rotation and boost
generators of Foldy [Eqs. (\ref{Rgenerator}) and (\ref{K})] are partitioned
differently into orbital and spin parts%
\begin{eqnarray}
\underline{{\bf J}} &=&-{\bf p\times }\underline{{\bf r}}_{{\rm P}}{\bf +%
\hat{p}}\,{\bf \hat{p}\cdot S}  \label{Skag} \\
\underline{{\bf K}} &=&\frac{1}{2}\left( \underline{{\bf r}}_{{\rm P}}p+p%
\underline{{\bf r}}_{{\rm P}}\right) .
\end{eqnarray}%
While components of \underline{${\bf r}$}$_{{\rm P}}$ do satisfy the
canonical commutation relations with ${\bf p,}$ they do not commute with
each other \cite{Skag94,Skagerstam}. Instead, they have a nonvanishing
commutator that is analogous to the field of a Dirac monopole (but in
momentum space) whose ``charge'' is given by the helicity operator ${\bf 
\hat{p}\cdot }$\underline{${\bf S}$} \cite{Mourad}.%
\begin{equation}
\left[ \underline{r}_{{\rm P}j},\underline{r}_{{\rm P}k}\right]
=-i\varepsilon _{jkl}\frac{p_{l}}{p^{3}}{\bf \hat{p}\cdot }\underline{{\bf S}%
}\,.  \label{S4.10}
\end{equation}%
As discussed in the next section, such monopole terms occur frequently in
commutators with gauge potentials.

\section{Photon position operators with commuting components}

In I, a modification of the Pryce position operator was constructed to have
commuting components . In this section we summarize the results of I,
emphasizing how the new position operator can be expressed as a spin
rotation of the weight-modified position operator for massive particles. We
then extend the construction to an arbitrary functional dependence $\chi _{%
{\bf p}}\left( \theta ,\phi \right) $ among the three Euler angles $\phi
,\theta ,\chi $.

The position operator 
\begin{equation}
\underline{{\bf r}}^{(0)}=ip^{\alpha }\nabla p^{-\alpha }+\frac{{\bf p\times 
}\underline{{\bf S}}}{p^{2}}-{\bf a}^{\left( 0\right) }\,{\bf \hat{p}\cdot }%
\underline{{\bf S}},  \label{rop}
\end{equation}%
with ${\bf a}^{\left( 0\right) }=\bbox{\hat{\phi }}\,p^{-1}\cot \theta \,,$
was constructed in Ref.\cite{Hawton1} by requiring that the components of 
\underline{${\bf r}$}$^{(0)}$ have transverse eigenvectors in the directions 
$\bbox{\hat{\theta }}$ and $\bbox{\hat{\phi }}$. The operator \underline{$%
{\bf r}$}$^{\left( 0\right) }$ can be expressed as a spin rotation of $%
ip^{\alpha }\nabla p^{-\alpha }:$%
\begin{equation}
\underline{{\bf r}}^{(0)}=i\underline{D}\left( \phi ,\theta ,0\right)
p^{\alpha }\nabla p^{-\alpha }\underline{D}^{-1}\left( \phi ,\theta
,0\right) \,.  \label{r0op}
\end{equation}%
Since this has the same form as (\ref{FSh}), it is equivalent to using $%
ip^{\alpha }\nabla p^{-\alpha }$ as the corresponding position operator in
the helicity representation. The construction (\ref{r0op}) may be understood
by analyzing the problems of the Pryce position operator (\ref{rP}). The
reason components of $r_{{\rm P}}$ do not commute is that the photon spin
and its momentum are inextricably coupled through the restriction on allowed
spin states, and in particular, the components of the momentum-space wave
function depend on momentum direction. Consequently, the action of the usual
momentum-space position operator for massive particles, $i\nabla ,$ is no
longer restricted to the argument of the wave function but also acts on its
components. In expression (\ref{r0op}), on the other hand, the spin rotation 
$D^{-1}$ rotates vector components of the state function from the lab to the
reference frame, where ${\bf \hat{p}}={\bf e}_{3},$ so that the operator $%
i\nabla $ can induce translations in its arguments without mixing its
components. Finally, $D$ rotates the translated vector back to the lab
frame. This is a straightforward way to restore the usual translational role
of $i\nabla ,$ and it is required to ensure that helicity eigenstates are
not mixed by infinitesimal momentum-space translations generated by 
\underline{${\bf r}$}$^{\left( 0\right) }$ away from the origin $p=0$. This
is an important property that follows from the invariance of the helicity
subspaces under the Poincar\'{e} group and has been confirmed, for example,
in experiments demonstrating Berry's phase \cite{Berry} for photons in
curved optical fibers \cite{Chiao86}.

The term ${\bf a}^{\left( 0\right) }=\bbox{\hat{\phi
}}\,p^{-1}\cot \theta $ multiplying the helicity operator ${\bf \hat{p}\cdot 
}$\underline{${\bf S}$} in the new position operator (\ref{rop}) is singular
in the limits $\theta \rightarrow 0,\pi ,$ that is, as ${\bf p}$ approaches
the $\pm {\bf e}_{3}$ axes, and has the momentum-space form of the
electromagnetic vector potential of a pair of Dirac monopole strings in
coordinate space. The relationship is discussed in more detail at the end of
Section V. It seems curious for the position operator to depend on the
choice of coordinates in the laboratory. The singularity together with the
dependence on laboratory coordinates may have deterred other authors from
including such a term. Nevertheless, we show below that some form of this
term is required to give the correct phase changes of the photon state under
rotations. Laboratory coordinates enter the formulation of \underline{${\bf r%
}$}$^{\left( 0\right) }$ in the spin rotation \underline{$D$} from ${\bf e}%
_{3}$ to ${\bf \hat{p}}$ and in the implicit dependence of ${\bf \hat{p}}$
on its polar angles $\theta ,\phi $. The singular term arises from the
differentiation of this rotation and from the path-dependent value of $\phi
. $ It is consistent with the relatively large changes in $\phi $ that can
result from infinitesimal changes in ${\bf p}$ at the string. In this
section we explore a generalization of the rotation used in \underline{${\bf %
r}$}$^{(0)}$ and show that the breakdown of the expected symmetry is related
to the selection of a specific ``gauge'' for \underline{${\bf r}$}${\bf \,.}$

The spin rotation that plays a central role in this paper can be written in
Cartesian form as the real unitary matrix of elements 
\begin{equation}
D_{jk}\equiv {\bf e}_{j}\cdot {\bf e}_{{\bf p}k}  \label{LE}
\end{equation}%
that relate vector components in the orthonormal lab basis $\left\{ {\bf e}%
_{1},{\bf e}_{2},{\bf e}_{3}\right\} $ to those in the photon-momentum basis 
$\left\{ {\bf e}_{{\bf p}1},{\bf e}_{{\bf p}2},{\bf e}_{{\bf p}3}\right\} ,$
where ${\bf e}_{{\bf p}3}\equiv {\bf \hat{p}.}$ Thus, for any vector ${\bf V,%
}$ $V_{j}={\bf V\cdot e}_{j}=D_{jk}{\bf V\cdot e}_{{\bf p}k}\equiv D_{jk}V_{%
{\bf p}k}\,.$ This gives the action of the matrix \underline{$D$} as a
passive transformation (${\bf V}$ is fixed, the ``observer'' basis changes).
However, only the relative orientation of the vector to the observer basis
enters the formalism, and in particular, ${\bf e}_{j}{\bf \cdot e}%
_{k}=\delta _{jk}={\bf e}_{{\bf p}j}{\bf \cdot e}_{{\bf p}k}\,.$ As a
result, an active interpretation of the action of \underline{$D$} (observer
fixed, ${\bf V}$ changes) is also possible and often more natural. As an
active transformation, \underline{$D$} rotates each of the vectors 
\underline{${\bf e}_{1}$}$,\underline{{\bf e}_{2}},$\underline{${\bf e}_{3}$}%
, of the lab frame into the corresponding vector \underline{${\bf e}_{{\bf p}%
1}$}$,\underline{{\bf e}_{{\bf p}2}},$\underline{${\bf e}_{{\bf p}3}$}, of
the photon ${\bf p}$-frame, with both vectors expressed in the lab basis: 
\underline{${\bf e}_{{\bf p}j}$}$=\underline{D}\,$\underline{${\bf e}_{j}$}%
\thinspace . The mathematical consistency of the different interpretations
is readily confirmed by writing out the column matrix representations of the
vectors. Further details are given in Appendix B.

The body rotation \underline{$D$} generalizes the polar rotation used in I.%
{\em \ }One can express \underline{$D$} in terms of spin matrices and the
Euler angles $\phi ,\theta $ and $\chi $ as in Eq.\thinspace (\ref{Drot}),
and it is given explicitly in Eq.\thinspace (\ref{Dmatrix}). The treatment
in I corresponds to the choice $\chi =0,$ in which ${\bf e}_{{\bf p}1}=%
\bbox{\hat{\theta}},$ ${\bf e}_{{\bf p}2}=\bbox{\hat{\phi}},$ and ${\bf e}_{%
{\bf p}3}={\bf \hat{p}\,.}$ For more general $\chi $, 
\begin{eqnarray}
{\bf e}_{{\bf p}1} &=&\bbox{\hat{\theta }}\cos \chi +\bbox{\hat{\phi }}\sin
\chi \text{,}  \label{ep1} \\
{\bf e}_{{\bf p}2} &=&\bbox{\hat{\phi }}\cos \chi -\bbox{\hat{\theta }}\sin
\chi .  \label{ep2}
\end{eqnarray}%
The angle $\chi $ represents an axial rotation about ${\bf \hat{p}}$ and is
associated below with the phase of a photon with definite helicity. As seen
by inspection of Eqs.\thinspace (\ref{ep1},\ref{ep2}), the transverse unit
vectors ${\bf e}_{{\bf p}1},{\bf e}_{{\bf p}2},$ are just the polar unit
vectors $\bbox{\hat{\theta},\hat{\phi},}$ rotated about ${\bf \hat{p}}$
through the angle $\chi $, so that use of the Euler angle $\chi $ allows an
arbitrary orientation of the two directions ${\bf e}_{{\bf p}1},{\bf e}_{%
{\bf p}2},$ in the transverse plane. The unit vectors describing states of
definite helicity $\kappa $, 
\begin{eqnarray}
{\bf e}_{{\bf p}\kappa } &\equiv &\frac{1}{\sqrt{2}}\left( {\bf e}_{{\bf p}%
1}+i\kappa {\bf e}_{{\bf p}2}\right)  \label{ehelicity} \\
&=&e^{-i\kappa \chi }{\bf e}_{{\bf p}\kappa }^{(0)},
\end{eqnarray}%
are just phase shifted relative to the helicity states ${\bf e}_{{\bf p}%
\kappa }^{(0)}=\left( \bbox{\hat{\theta }}+i\kappa \bbox{\hat{\phi }}\right)
/\sqrt{2}$ with $\chi =0$. Under the general rotation (\ref{Drot}), the spin
matrices themselves are transformed as 
\begin{equation}
\underline{S_{{\bf p}j}}=\underline{D}\,\underline{S_{j}}\,\underline{D}%
^{-1}=\underline{S_{k}}D_{kj}\text{,}  \label{Sp}
\end{equation}%
where in particular \underline{$S_{{\bf p}3}$}$=\underline{{\bf S}}{\bf %
\cdot \hat{p}\,.}$ Consequently, we can also express \underline{$D$} in
terms of rotations about the ${\bf p}$-frame axes: 
\begin{eqnarray}
\underline{D} &=&\underline{D}\,\underline{D}\,\underline{D}^{-1} \\
&=&e^{-i\underline{D}\underline{S_{3}}\underline{D}^{-1}\phi }e^{-i%
\underline{D}\underline{S_{2}}\underline{D}^{-1}\theta }e^{-i\underline{D}%
\underline{S_{3}}\underline{D}^{-1}\chi }  \nonumber \\
&=&\exp \left( -i\underline{S_{{\bf p}3}}\phi \right) \exp \left( -i%
\underline{S_{{\bf p}2}}\theta \right) \exp \left( -i\underline{S_{{\bf p}3}}%
\chi \right) .
\end{eqnarray}%
Alternatively, we can make the axial rotation the last of the Euler-angle
rotations by putting \underline{$D$}$_{0}\equiv \exp \left( -i\underline{%
S_{3}}\phi \right) \exp \left( -i\underline{S_{2}}\theta \right) $ and noting%
\begin{eqnarray}
\underline{D} &=&\underline{D}_{0}\exp \left( -i\underline{S_{3}}\chi
\right) \underline{D}_{0}^{-1}\underline{D}_{0} \\
&=&\exp \left( -i\underline{D}\underline{S_{3}}\underline{D}^{-1}\chi
\right) \underline{D}_{0} \\
&=&\exp \left( -i\underline{{\bf S}}{\bf \cdot \hat{p}\,}\chi \right) \exp
\left( -i\underline{S_{3}}\phi \right) \exp \left( -i\underline{S_{2}}\theta
\right) .
\end{eqnarray}

The position operator that generalizes Eq.\thinspace (\ref{rop}) is the $%
3\times 3$ matrix 
\begin{eqnarray}
\underline{{\bf r}} &=&\underline{D}\left( ip^{\alpha }\nabla p^{-\alpha
}\right) \underline{D}^{-1}  \label{rralph} \\
&=&ip^{\alpha }\nabla p^{-\alpha }-\underline{{\bf A}}\,,  \label{raop}
\end{eqnarray}%
where 
\begin{eqnarray}
\underline{{\bf A}} &=&-i\underline{D}\left( \nabla \underline{D}%
^{-1}\right) =i\left( \nabla \underline{D}\right) \underline{D}^{-1} \\
&=&\frac{\underline{{\bf S}}\times {\bf \hat{p}}}{p}+{\bf a\,\hat{p}\cdot }%
\underline{{\bf S}}\,,
\end{eqnarray}%
and ${\bf a}$ is a vector with dimensions of length. If $\chi $ is
independent of $\theta ,\phi ,$ then the axial rotation elements commute
with $\nabla $ and $\underline{{\bf r}}$ is given by $\underline{{\bf r}}%
^{\left( 0\right) }$ as found in I. However, the momentum ${\bf p}$ is
invariant under the axial rotation by $\chi ,$ and we can choose a
functional dependence $\chi =\chi _{{\bf p}}\left( \theta ,\phi \right) .$
Then the matrix \underline{$D$} is a function of the two parameters $\theta
,\phi ,$ and direct calculation of the gradient of the rotation matrix gives%
\begin{equation}
{\bf a}={\bf a}^{\left( 0\right) }+\nabla \chi _{{\bf p}}  \label{a}
\end{equation}%
with%
\begin{equation}
{\bf a}^{\left( 0\right) }=\bbox{\hat{\phi }}\frac{\cot \theta }{p}=\frac{%
{\bf e}_{3}\times {\bf p\,\hat{p}\cdot e}_{3}}{\left( {\bf e}_{3}\times {\bf %
p}\right) ^{2}}.  \label{a0}
\end{equation}%
When operating on a subspace of definite helicity $\kappa =\pm 1,$ 
\underline{${\bf r}$} is related to the Pryce operator $\underline{{\bf r}}_{%
{\rm P}}$ of form (\ref{rP}) by a transverse displacement whose sign depends
on $\kappa $ :%
\begin{equation}
\underline{{\bf r}}=\underline{{\bf r}}_{{\rm P}}-\kappa {\bf a\,.}
\label{rrP}
\end{equation}%
This clearly manifests the basic coupling between the helicity (and
therefore the spin) and the position. However, as we see in the following
section, the eigenvalues of \underline{${\bf r}$} do not depend explicitly
on $\kappa \,.$

The matrix \underline{${\bf A}$}, with transverse vectors as its matrix
elements, has a simple interpretation. Writing%
\begin{equation}
\nabla \underline{D}=-i\underline{{\bf A}}\,\underline{D}
\end{equation}%
and making the dependence of the rotation matrix \underline{$D$} on ${\bf p}$
explicit, we find%
\begin{eqnarray}
\underline{D}\left( {\bf p}+d{\bf p}\right) &=&\underline{D}\left( {\bf p}%
\right) +d{\bf p\cdot \nabla }\underline{D}\left( {\bf p}\right) \\
&=&\left( 1-id{\bf p\cdot }\underline{{\bf A}}\right) \underline{D}\left( 
{\bf p}\right)  \label{Dpdp}
\end{eqnarray}%
Thus, \underline{${\bf A}$}$\cdot d{\bf p}$ generates the infinitesimal
rotation induced when ${\bf p}$ is incremented by $d{\bf p}$.

The possibility of choosing different functions $\chi _{{\bf p}}\left(
\theta ,\phi \right) $ dependent on the position $\theta ,\phi $ in
parameter space expresses the invariance of ${\bf p}$ under axial rotations
as a local symmetry and is analogous to gauging ${\bf a}$ and hence 
\underline{${\bf r}$}${\bf .}$ The gauge group of axial rotations is $U(1),$
and ${\bf a}$ is thus analogous to the electromagnetic vector potential. The
scalar gauging function $\chi _{{\bf p}}$ is a concrete example of Berry's
gauge \cite{Berry} that sets the local phase of an eigenstate in parameter
space. It also appears closely related to the function $G_{0}$ introduced by
Staruszkiewicz in his earlier study \cite{Starus73} of parallel transport of
photons. For all choices of single-valued differentiable functions $\chi _{%
{\bf p}}$, the gauge field%
\begin{equation}
\nabla \times {\bf a}=-\frac{{\bf \hat{p}}}{p^{2}},\quad \sin \theta \neq 0,
\label{curla}
\end{equation}%
has the same monopole form. It follows that ${\bf a}$ itself cannot be
replaced by a gradient term. As we discuss below (see especially Section V)
however, $\chi _{{\bf p}}$ is generally not single valued and can affect
stringlike singularities in the field (\ref{curla}). The ``abelian vector
potential'' ${\bf a}$ is just part of a nonabelian operator $\underline{{\bf %
A}}=\underline{{\bf S}}\times {\bf \hat{p}}/p+{\bf a\,\hat{p}\cdot }%
\underline{{\bf S}}\,.$

Berry \cite{Berry} has found analogous monopole fields (\ref{curla}) both
for the case of a pair of adiabatic states that are degenerate at an
isolated point in parameter space and for spin-$S$ charges in a magnetic
field that are degenerate where the field vanishes. Our derivation shows
that such fields arise more generally in calculations of topological phase,
even when there is no isolated point of degeneracy. Indeed, monopole terms
of the form ${\bf \hat{p}}/p^{2}$ frequently arise in the commutators
involving gauge potentials associated with massless representations of the
Poincar\'{e} group \cite%
{Mourad,BialynickiBirula,Skag94,Skagerstam,Bala83,Wu75,Godd78,Pati96}. In
our case, direct calculation gives%
\begin{eqnarray}
\left[ \underline{A}_{j},\underline{A}_{k}\right] &=&i\varepsilon
_{jkl}\left( \nabla _{{\bf p}}\times \underline{{\bf A}}\right) \cdot {\bf e}%
_{l}\, \\
&=&i\varepsilon _{jkl}\left( {\bf \hat{p}\cdot }\underline{{\bf S}}\frac{%
{\bf a\times p}+{\bf \hat{p}}}{p^{2}}+\frac{\underline{{\bf S}}{\bf \times a}%
}{p}\right) \cdot {\bf e}_{l}\,,
\end{eqnarray}%
which includes the monopole term ${\bf \hat{p}}/p^{2}$ multiplied by the
helicity operator ${\bf \hat{p}\cdot \underline{S}.}$ Similar terms also
arise from the distinct ``covariant derivatives'' ${\frak D}^{\prime
}=\nabla -i{\bf p\times S}/p^{2}$ and ${\frak D}^{\prime \prime }=\nabla
-i\kappa {\bf a}$ introduced for photon operators by Pati \cite{Pati96} and
the Bia\l ynicki-Birulas \cite{BialynickiBirula}, respectively. The common
appearance of monopole terms and their associated string singularity (see
Section V) is a direct consequence of the topology of momentum space \cite%
{Starus73,Skag87} (see also Appendix A). Here, the monopole ``field'' (\ref%
{curla}) turns out to be just what is required to cancel the singularity in
the commutator (\ref{S4.10}) of the Pryce position operators. Furthermore,
the relation $\left( \nabla +i\underline{{\bf A}}\right) \underline{D}=0$
implies that ${\frak D}=\nabla +i$\underline{${\bf A}$} is the appropriate
``covariant derivative'' in our formulation.

There is a wide range of possible gauge choices since $\chi _{{\bf p}}\left(
\theta ,\phi \right) $ can in principle be any differentiable function.
However, three choices of $\chi _{{\bf p}}$ are of particular interest: If $%
\chi _{{\bf p}}$ is the zero function, $\chi _{{\bf p}}^{\left( 0\right) }=0$%
, then the transverse directions are the usual spherical polar unit vectors, 
${\bf a}$ is just ${\bf a}^{\left( 0\right) }$ (\ref{a0}), and $\underline{%
{\bf r}}$ is given by Eq.(\ref{rop}). If $\chi _{{\bf p}}$ is $\chi _{{\bf p}%
}^{\left( 1\right) }=-\phi $, then as noted in Section II, \underline{$D$}
is the Chakrabarti \cite{Chakrabarti} transformation \underline{$U$} (\ref%
{Uh}), and ${\bf a}$ takes the form%
\begin{equation}
{\bf a}^{\left( 1\right) }=\frac{\cos \theta -1}{p\sin \theta }%
\bbox{\hat{\phi }}=\frac{{\bf \hat{p}\times e}_{3}}{p+{\bf p\cdot e}_{3}}
\label{fhelicity}
\end{equation}%
in Eq.(\ref{raop}). We note that this choice eliminates the singularity in $%
{\bf a}$ at $\theta =0,$ but doubles the strength of the one at $\theta =\pi 
$ (see also Section V). If instead $\chi _{{\bf p}}$ is $\chi _{{\bf p}%
}^{\left( 2\right) }=-\phi \cos \theta $, then ${\bf a}$ is given by 
\begin{equation}
{\bf a}^{\left( 2\right) }=\frac{\bbox{\hat{\theta }}}{p}\phi \sin \theta 
\text{.}  \label{a2}
\end{equation}%
In this case, both string singularities are removed, but ${\bf a}$ is now
path dependent, since the value of $\phi $ depends on how many times the
path wraps around the ${\bf e}_{3}$ axis. Consequently, ${\bf a}^{\left(
2\right) }$ is ``nonintegrable'' \cite{Wu75}. By subtracting another
path-dependent term, it is also possible to reintroduce a string singularity
about a different axis.

\section{Poincar\'{e} generators and commutation relations}

Here we examine the position operator \underline{${\bf r}$} in the context
of the Poincar\'{e} algebra and write explicit momentum-space expressions
for the localized bases. To see the relationship of our position operator 
\underline{${\bf r}$} and the Poincar\'{e} operators in the existing
literature, note that translation generators ${\bf p}$ and $H=pc$ are
unchanged, and the Foldy \underline{${\bf J}$} and \underline{${\bf K}$}
operators can be written in terms of an orbital angular momentum operator 
\begin{equation}
\underline{{\bf L}}^{\left( r\right) }\equiv \underline{{\bf r}}\times {\bf p%
}  \label{L}
\end{equation}%
corresponding to our position operator (\ref{raop}), and an effective spin
operator, which has a transverse contribution in addition to the helicity
term that is frequently assumed to be the total spin contribution\thinspace %
\cite{Skag94,Skagerstam,Jordan2}, 
\begin{equation}
\underline{{\bf S}}^{\left( r\right) }\equiv \left( {\bf a\times p}+{\bf 
\hat{p}}\right) {\bf \hat{p}\cdot \underline{S}\,.}  \label{Sr}
\end{equation}%
We obtain 
\begin{equation}
\underline{{\bf J}}=\underline{{\bf L}}^{\left( r\right) }+\underline{{\bf S}%
}^{\left( r\right) }  \label{Jr}
\end{equation}%
and 
\begin{eqnarray}
\underline{{\bf K}} &=&\frac{1}{2}\left( p\underline{{\bf r}}{\bf +}%
\underline{{\bf r}}p\right) +{\bf \hat{p}}\times \underline{{\bf S}}^{\left(
r\right) }  \label{Kr} \\
&=&\frac{1}{2}\left( p\underline{{\bf r}}{\bf +}\underline{{\bf r}}p\right) +%
{\bf a\,p\cdot \underline{S}}
\end{eqnarray}%
as the Foldy operators (generalized to arbitrary $\alpha $ ) in terms of our
position operator. With these expressions, transformation to one of the
helicity forms is straightforward. By inverting Eqs.(\ref{rralph}) and (\ref%
{Sp}) the position operator becomes \underline{$D$}$^{-1}\underline{{\bf r}}%
{\bf \,}\underline{D}=ip^{\alpha }\nabla p^{-\alpha }$, while \underline{$D$}%
$^{-1}{\bf \hat{p}\cdot \underline{S}}\,\underline{D}=$\underline{$S_{3}$}.
The general helicity representation is given by \underline{$O$}$_{{\rm h}}=%
\underline{D}^{-1}\underline{O}\underline{D},$ which gives%
\begin{eqnarray}
\underline{{\bf J}}_{{\rm h}} &=&\left( ip^{\alpha }\nabla p^{-\alpha
}\right) \times {\bf p+}\left( {\bf a\times p}+{\bf \hat{p}}\right) 
\underline{S_{3}} \\
\underline{{\bf K}}_{{\rm h}} &=&\frac{i}{2}p^{\alpha }\left( p{\bf \nabla
+\nabla }p\right) p^{-\alpha }+{\bf a}p\underline{S_{3}}
\end{eqnarray}%
If $\chi _{{\bf p}}=0$, then ${\bf a}=\bbox{\hat{\phi}}\cot \theta /p$ and 
\underline{${\bf J}$}$_{{\rm h}}$ and \underline{${\bf K}$}$_{{\rm h}}$
reduce to the Shirokov operators given by Eqs.(\ref{JS}) and (\ref{KS})
generalized to arbitrary $\alpha $, while if $\chi _{{\bf p}}=-\phi $, ${\bf %
a}$ is given by $\bbox{\hat{\phi }}\left( \cos \theta -1\right) /\left(
p\sin \theta \right) =-\bbox{\hat{\phi}}\sin \theta /\left( 1+\cos \theta
\right) $ and \underline{${\bf J}$}$_{{\rm h}}$ and \underline{${\bf K}$}$_{%
{\rm h}}$ reduce to the Lomont and Moses operators, Eqs.(\ref{Jh}) and (\ref%
{Kh}) with arbitrary $\alpha $. Thus both the Lomont and Moses and the
Shirokov transformations are special cases of Euler angle rotations relative
to the Foldy form. In any irreducible helicity representation the position
operator is just $ip^{\alpha }\nabla p^{-\alpha }\ $as it is for massive
particles in the Foldy form, since the transverse directions are fixed in
the reference frame where the photon momentum is parallel to ${\bf e}_{3}$.
The helicity reference frame is the photon frame, while the Foldy frame is
the lab frame.

The commutation relations satisfied by the generators of time and space
translations, rotations and boosts, $H=pc$, $p_{i}$, \underline{$J_{j}$} and 
\underline{$K_{j}$}, are the standard ones required by the Poincar\'{e}
algebra: $\left[ \underline{J_{j}},\underline{J_{k}}\right] =i\epsilon
_{jkl} $\underline{$J_{l}$}, $\left[ \underline{J_{j}},\underline{K_{k}}%
\right] =i\epsilon _{jkl}$\underline{$K_{l}$}, $\left[ \underline{K_{j}},%
\underline{K_{k}}\right] =-i\epsilon _{jkl}$\underline{$J_{l}$}, $\left[ 
\underline{J_{j}},p_{k}\right] =i\epsilon _{jkl}p_{l}$, $\left[ \underline{%
K_{j}},p_{k}\right] =i\delta _{jk}H/c$, $\left[ \underline{K_{j}},H\right]
=icp_{j}$, $\left[ \underline{J_{j}},H\right] =0$, $\left[ p_{j},H\right] =0$%
. Analogous commutation relations are valid if \underline{$L_{j}$}$^{\left(
r\right) }$ is substituted for \underline{$J_{j}$} and $\left( p\underline{%
r_{j}}+\underline{r_{j}}p\right) /2$ for \underline{$K_{j}$}, since they are
unitarily equivalent to the zero-spin case of the above. Commutation
relations involving \underline{$r_{j}$} in the lab frame can be most simply
derived by performing a matrix rotation to the photon reference frame, in
which \underline{${\bf r}$} is replaced by $ip^{\alpha }\nabla p^{-\alpha }$
and noting that components $p_{j}$ in the momentum representation commute
with the rotation matrix \underline{$D$}$.$ Thus $\left[ \underline{r_{j}},%
\underline{r_{k}}\right] =0$, $\left[ \underline{r_{j}},p_{k}\right]
=i\delta _{jk}$, $\left[ \underline{r_{j}},H\right] =iHp_{j}/p^{2}$, $\left[ 
\underline{L_{j}}^{\left( r\right) },\underline{r_{k}}\right] =i\epsilon
_{jkl}$\underline{$r_{l}$}, $\left[ \underline{r_{j}},\underline{S_{{\bf p}k}%
}\right] =0$ and $\left[ \underline{L_{j}}^{\left( r\right) },\underline{S_{%
{\bf p}k}}\right] =0$. It follows from the last two relations that a photon
can simultaneously have definite helicity and either a definite position or
a well-defined spatial-component of orbital angular momentum. In other
words, measurement of either the position or the orbital angular momentum of
a photon state does not change its helicity. Furthermore, because 
\begin{equation}
\left[ \underline{J_{k}},{\bf \hat{p}\cdot S}\right] =0,\;\left[ \underline{K%
}_{k},{\bf \hat{p}\cdot S}\right] =0,\;\left[ p_{k},{\bf \hat{p}\cdot S}%
\right] =0,
\end{equation}%
the helicity is an invariant of the Poincar\'{e} group; it is invariant
under all rotations, boosts, and translations (as long as these avoid the
origin $p=0$). Consequently, every representation of the Poincar\'{e} group
for photons can be reduced to the direct sum of representations for the two
helicities, and every irreducible representation will be carried by states
of a single helicity. It may therefore be convenient to specify the
operators for the separate invariant helicity subspaces. The momentum and
energy operators are unchanged, the position operator is as given above in
Eq.\thinspace (\ref{rrP}), and \underline{${\bf J}$} and \underline{${\bf K}$%
} take exactly their Pryce forms 
\begin{equation}
\underline{{\bf J}}=\underline{{\bf L}}^{\left( r\right) }+\underline{{\bf S}%
}^{\left( r\right) }=\underline{{\bf r}}_{{\rm P}}\times {\bf p+\hat{p}}%
\kappa
\end{equation}%
and 
\begin{equation}
\underline{{\bf K}}=\frac{1}{2}\left( p\underline{{\bf r}}{\bf +}\underline{%
{\bf r}}p\right) +{\bf a\,}\kappa =\frac{1}{2}\left( p\underline{{\bf r}}_{%
{\rm P}}+\underline{{\bf r}}_{{\rm P}}p\right)
\end{equation}%
with \underline{${\bf r}$}$_{{\rm P}}=\underline{{\bf r}}+\kappa {\bf a\,.}$

It is important that our position operators obey the correct dynamical
equations. In the Heisenberg picture, the dynamics are determined by the
equation of motion, which from Eq.\thinspace (\ref{rralph}) is 
\begin{equation}
\frac{d\underline{{\bf r}}}{dt}=\frac{\partial \underline{{\bf r}}}{\partial
t}+i\left[ H,\underline{{\bf r}}\right] =i[cp,\underline{{\bf r}}]=c{\bf 
\hat{p}}\,.
\end{equation}%
Thus our theory predicts that the photon has a velocity ${\bf v}=c{\bf \hat{p%
}}$, as required. We also note that \underline{${\bf r}$} (\ref{raop}) is
Hermitian and symmetric under time reversal. One may in addition expect 
\underline{${\bf r}$} to change sign under parity inversion, but this
depends on the gauge potential. It is valid for \underline{${\bf r}$}$%
^{\left( 0\right) },$ but for other gauges we must generally replace the
inversion in $\chi _{{\bf p}}$ by the gauge change $\chi _{{\bf p}%
}\rightarrow -\chi _{{\bf p}}$ in order to ensure the invariance of $\nabla
\chi _{{\bf p}}\,.$

In general, the position operator rotates as a simple vector under 
\underline{${\bf L}$}$^{\left( r\right) }$ since $\left[ \underline{L_{j}}%
^{\left( r\right) },\underline{r_{k}}\right] =i\epsilon _{jkl}\underline{%
r_{l}}\,,$ but not under \underline{${\bf J}$} :%
\begin{equation}
\left[ \underline{J_{j}},\underline{r_{k}}\right] =i\epsilon _{jkl}%
\underline{r_{l}}+\left[ \underline{S_{j}}^{\left( r\right) },\underline{%
r_{k}}\right] \,.  \label{comm}
\end{equation}%
From $\left[ {\bf \hat{p}\cdot }\underline{{\bf S}},\underline{r_{j}}\right]
=0$ and Eq.(\ref{Sr}), we thus find%
\begin{eqnarray}
\left[ \underline{J_{j}},\underline{r_{k}}\right] &=&i\epsilon _{jkl}%
\underline{r_{l}}+\left[ \left( {\bf a\times p+\hat{p}}\right) \cdot {\bf e}%
_{j},\underline{r_{k}}\right] \kappa  \label{rJcomm} \\
&=&i\epsilon _{jkl}\underline{r_{l}}-i\kappa \left\{ \frac{\partial }{%
\partial p_{k}}\left( {\bf a\times p+\hat{p}}\right) \cdot {\bf e}%
_{j}\right\} \text{.}  \nonumber
\end{eqnarray}%
The extra term represents a deviation from the usual commutator (\ref{commJ}%
), and is due to the coupling of the momentum and spin of a photon. Note
that the deviation vanishes for rotations about ${\bf \hat{p}.}$ It also
vanishes in the special case of a rotation about ${\bf e}_{3}$ when $\chi _{%
{\bf p}}$ is independent of $\phi $.

For a photon with helicity $\kappa $ the momentum-space states localized at $%
{\bf r}$ are 
\begin{equation}
\underline{{\bf \Psi }_{{\bf r}^{\prime },\kappa }}\left( {\bf p}\right)
=Np^{\alpha }e^{-i{\bf r}^{\prime }\cdot {\bf p}}\underline{{\bf e}_{{\bf p}%
\kappa }}\text{,}  \label{localized}
\end{equation}%
as may be verified by direct application of the operator ${\bf r}$ (\ref%
{rralph}). In particular, the localized state at the origin,%
\begin{equation}
\underline{{\bf \Psi }_{0,\kappa }}\left( {\bf p}\right) =Np^{\alpha }%
\underline{{\bf e}_{{\bf p}\kappa }}  \label{local0}
\end{equation}%
gives%
\begin{equation}
\underline{{\bf L}}^{(r)}\underline{{\bf e}_{{\bf p}\kappa }}=-{\bf p}\times 
\underline{{\bf r}}\underline{{\bf e}_{{\bf p}\kappa }}=0
\end{equation}%
since $\underline{{\bf \Psi }_{0,\kappa }}$ is an eigenvector of \underline{$%
{\bf r}$} with eigenvalue $0$. The relation \underline{${\bf r}$}$\,$%
\underline{${\bf e}_{{\bf p}\kappa }$}$=0$ also follows directly from the
transformations (\ref{rralph}) and (\ref{epk}). Using \underline{${\bf J}$}$=%
\underline{{\bf L}}^{(r)}+\underline{{\bf S}}^{(r)}$ where \underline{${\bf S%
}$}$^{(r)}$\ is given by Eq.(\ref{Sr}), we find the rotation by an
infinitesimal angle $d\xi $ about the axis $\bbox{\hat{\xi}}$\ to be 
\begin{eqnarray}
\exp \left( -i\underline{{\bf J}}\cdot d\bbox{\xi }\right) \underline{{\bf e}%
_{{\bf p}\kappa }} &=&\exp \left( -i\underline{{\bf S}}^{(r)}\cdot d%
\bbox{\xi }\right) \underline{{\bf e}_{{\bf p}\kappa }} \\
&=&\exp \left( -i\kappa ({\bf a\times p+\hat{p}})\cdot d\bbox{\xi }\right) 
\underline{{\bf e}_{{\bf p}\kappa }}  \nonumber
\end{eqnarray}%
with $d\bbox{\xi }=\bbox{\hat{\xi}}d\xi $ . Thus the rotated transverse
basis vectors of helicity $\kappa $ satisfy%
\begin{equation}
\exp \left( -i\underline{{\bf J}}\cdot d\bbox{\xi }\right) \underline{{\bf e}%
_{{\bf p}\kappa }}=\exp \left[ -i\kappa \left( d\chi -d\chi _{{\bf p}%
}\right) \right] \underline{{\bf e}_{{\bf p}\kappa }},  \label{roteps}
\end{equation}%
and are thus changed only by an infinitesimal phase shift%
\begin{equation}
-\kappa \left( d\chi -d\chi _{{\bf P}}\right) =-\kappa ({\bf a}\times {\bf p+%
}{\bf \hat{p}})\cdot d\bbox{\xi }\text{.}  \label{delta-chi}
\end{equation}%
Here, $d\chi _{{\bf p}}=d{\bf p}\cdot \nabla \chi _{{\bf p}}$ is the change
in the function $\chi _{{\bf p}}\left( \theta ,\phi \right) $ that results
from the change $d{\bf p}$ in the photon momentum. (Note that ${\bf a}$ and $%
\underline{{\bf S}}^{(r)}$ depend on the orientation of ${\bf \hat{p},}$ and
therefore ordered integral expressions of the Dyson type are required for
finite rotations. See also the Appendix.)

Since ${\bf e}_{{\bf p}0}\equiv {\bf \hat{p}}$ is a longitudinal basis
vector corresponding to helicity $\kappa =0$, it follows from ${\bf \hat{p}%
\cdot }\underline{{\bf S}}\,\underline{{\bf e}_{{\bf p}0}}=0$ that
Eq.\thinspace (\ref{roteps}) is also true of the longitudinal vector.
Indeed, Eq.\thinspace (\ref{roteps}) may be considered an extension of the
expected invariance of the radial vector field ${\bf \hat{p}}$ under
rotations. Also, if $\bbox{\hat{\xi}=e}_{3}$\thinspace , and $\chi _{{\bf p}%
}\ $is constant, then it is readily shown that $d\chi -d\chi _{{\bf P}}=0$
so that this cylindrically symmetric transverse vector field is invariant
under rotations about ${\bf e}_{3}$. Note further from Eq.\thinspace (\ref%
{Dpdp}) that with $d{\bf p}=d\bbox{\xi }\times {\bf p,}$ the infinitesimal
rotation factor becomes%
\begin{equation}
\underline{D}\left( {\bf p}+d{\bf p}\right) \underline{D}^{-1}\left( {\bf p}%
\right) =1-i\underline{{\bf A}}\cdot d{\bf p}=1-i\left( {\bf S}-{\bf S}%
^{\left( r\right) }\right) \cdot d\bbox{\xi \,.}
\end{equation}%
The part $-i{\bf S}\cdot d\bbox{\xi }$ gives the additional rotation whereas 
$i{\bf S}^{\left( r\right) }\cdot d\bbox{\xi }$ corrects the axial rotation
implied by the functional dependence $\chi _{{\bf p}}\left( \theta ,\phi
\right) .$

One can understand the extra term in the commutator of the position with the
angular momentum by comparing Eqs.(\ref{rJcomm}) and (\ref{delta-chi}).
Rotations generally change the ``gauge potential'' ${\bf a}$ in a way
analogous to transformations of the electromagnetic potential $A$ under
Lorentz transformations. The $\partial \left( {\bf a\times p+\hat{p}}\right)
\cdot {\bf e}_{j}/\partial p_{i}$ term in Eq.(\ref{rJcomm}) is required in
order to give the axial spin rotation needed by the matrices associated with
each component of \underline{${\bf r}$}\thinspace . Such axial rotations
result in the phase change of eigenvectors during a rotation that maintains
the functional dependence of $\chi _{{\bf p}}$ and is required in order to
give the correct Berry's phase for photons (see the next section). The
possible transformations are discussed in more detail in the Appendices, and
Eqs.(\ref{delta-chi}) and (\ref{rJcomm}) are independently derived in
Appendices B and C respectively.

\section{Relation to Berry's phase}

The phase angle $-\kappa \left( d\chi -d\chi _{{\bf p}}\right) $ is
important and has measurable consequences. In particular, it can be
integrated to give the total phase change of a state vector of helicity $%
\kappa $ when transported along a closed loop in parameter space. The result
depends on the type of transport, but it will not depend on any
single-valued functional choice of $\chi _{{\bf p}}.$ However, it is natural
to make $\chi _{{\bf p}}$ multi-valued and path-dependent (nonintegrable),
and in that case the phase change will depend linearly on the change $\Delta
\chi _{{\bf p}}$ around the closed loop, as we show explicitly below. If we
use parallel transport, the angle $-\kappa \left( d\chi -d\chi _{{\bf p}%
}\right) $ is Berry's phase \cite{Berry,Aharon} $\gamma _{\kappa }$, a
topological phase accumulated by the photon. This is consistent with a more
general association of Berry's phase in quantum systems with geometrical
angles.\cite{Anan}

While Berry's derivation assumed adiabatic transport of energetically
discrete states, his results can also be applied to degenerate helicity
states in parallel transport \cite{Anan,Chiao86}. Parallel transport on the
spherical surface of constant radius $p$ is most easily realized by piecing
the path together from many small segments of great circles and employing
nonrotating transport along each segment. Each great-circle segment requires
an axis of rotation that is perpendicular to ${\bf p}$ but generally changes
as ${\bf p}$ moves. From Eqs.\thinspace (\ref{delta-chi}) and (\ref{roteps}%
), the accumulated phase for great-circle segment (${\bf \hat{p}}\cdot d%
\bbox{\xi }=0$) is%
\begin{eqnarray}
-\kappa \left( \Delta \chi -\Delta \chi _{{\bf p}}\right) &=&-\kappa \oint 
{\bf a}\times {\bf p}\cdot d\bbox{\xi }  \nonumber \\
&{\bf =}&-\kappa \oint {\bf a}\cdot {\bf p}\times d\bbox{\xi }  \nonumber \\
&=&\kappa \oint {\bf a\cdot }d{\bf p\,,}  \label{lineint}
\end{eqnarray}%
where we noted $d{\bf p}=d\bbox{\xi \times p\,.}$ The same result (\ref%
{lineint}) can be obtained directly from Berry's derivation \cite{Berry} of
the geometric phase\thinspace $\gamma _{\kappa }$. We can write his starting
point in differential form as%
\begin{equation}
d\gamma _{\kappa }=i\left\langle u_{\kappa }\left( {\bf p}\right) |\nabla
u_{\kappa }\left( {\bf p}\right) \right\rangle \cdot d{\bf p}
\label{Berryeq4}
\end{equation}%
since the relevant parameter for the photon as it is guided in an optical
fiber is its momentum ${\bf p\,.}$ In our case, the eigenstate $u_{\kappa
}\left( {\bf p}\right) $ is proportional to the column vector%
\begin{equation}
\underline{{\bf e}_{{\bf p}\kappa }}=\underline{D}\,\underline{{\bf e}%
_{\kappa }}
\end{equation}%
so that Berry's relation (\ref{Berryeq4}) reduces to%
\begin{eqnarray}
d\gamma _{\kappa } &=&i\left\langle \underline{{\bf e}_{\kappa }}^{\dag }%
\underline{D}^{\dag }\left( \nabla \underline{D}\right) \,\underline{{\bf e}%
_{\kappa }}\right\rangle \cdot d{\bf p}=\left\langle \underline{{\bf e}%
_{\kappa }}^{\dag }\underline{D}^{\dag }\underline{{\bf A}}\,\underline{D}%
\underline{{\bf e}_{\kappa }}\right\rangle \cdot d{\bf p} \\
&=&\left\langle \underline{{\bf e}_{{\bf p}\kappa }}^{\dag }\underline{{\bf A%
}}\,\underline{{\bf e}_{{\bf p}\kappa }}\right\rangle \cdot d{\bf p}=\kappa 
{\bf a}\cdot d{\bf p}
\end{eqnarray}%
where the matrix sandwich indicated by angular brackets $\left\langle \cdots
\right\rangle $ is a vector-valued $1\times 1$ matrix, and we have noted
that \underline{${\bf S}$}$\times {\bf \hat{p}}$ has a vanishing diagonal in
the \underline{${\bf e}_{{\bf p}\kappa }$} basis. In terms of the polar
angles $\theta ,\phi ,$ displacement on the spherical surface is given by $d%
{\bf p}=\bbox{\hat{\phi}\,}p\sin \theta \,d\phi +\bbox{\hat{\theta}}pd\theta
\,,$ and therefore with Eqs.\thinspace (\ref{a}, \ref{a0}) for ${\bf a}$%
\thinspace , the line integral (\ref{lineint}) gives%
\begin{eqnarray}
-\kappa \left( \Delta \chi -\Delta \chi _{{\bf p}}\right) &=&\kappa \left(
\oint \cos \theta \,d\phi +\oint \nabla \chi _{{\bf p}}\cdot d{\bf p}\right)
\nonumber \\
&=&\kappa \left( \oint \cos \theta \,d\phi +\Delta \chi _{{\bf p}}\right) \,.
\label{Berrysphase}
\end{eqnarray}

The result depends on the function $\chi _{{\bf p}}\,.$ We must choose a
function that eliminates superfluous axial rotations. Recall that the basis
vectors \underline{${\bf e}_{{\bf p}\kappa }$} are obtained from the
reference frame by the rotation \underline{$D$}$\left( \phi ,\theta ,\chi _{%
{\bf p}}\right) $\underline{${\bf e}_{\kappa }$}. To avoid unwanted axial
rotations, we ensure that ${\bf p}$ is also parallel transported by this
rotation. This requires the function choice%
\begin{equation}
\chi _{{\bf p}}\left( \theta ,\phi \right) =\chi _{{\bf p}}^{\left( 1\right)
}\left( \theta ,\phi \right) =-\phi
\end{equation}%
to make \underline{$D$}$\left( \phi ,\theta ,\chi _{{\bf p}}\right) $
equivalent to the direct great-circle rotation from ${\bf e}_{3}$ to ${\bf 
\hat{p}.}$ The phase shift is then%
\begin{equation}
\gamma _{\kappa }=-\kappa \left( \Delta \chi -\Delta \chi _{{\bf p}}^{\left(
1\right) }\right) =\kappa \left( \oint \cos \theta \,d\phi -2\pi \right)
=-\kappa \Omega ,
\end{equation}%
where $\Omega $ is the solid angle enclosed by the loop. This is Berry's
geometrical phase \cite{Berry} which has been confirmed in experiments on
light in helically wound optical fibers \cite{Chiao86}. The sign convention %
\cite{Berry} is taken such that the dynamic phase of a stationary state (due
to a factor $\exp \left( -i\omega t\right) $) decreases in time (see also
Appendix D).

One often seeks to generalize the above result by an application of Stokes
theorem. In our case%
\begin{eqnarray}
\oint {\bf a}^{\left( 1\right) }{\bf \cdot }d{\bf p} &=&\int_{\Omega }\left( 
{\bf \nabla \times a}^{\left( 1\right) }\right) \cdot {\bf \hat{p}}%
p^{2}d\Omega \quad \quad  \nonumber \\
&=&-\int_{\Omega }{\bf \hat{p}}\cdot \frac{{\bf \hat{p}}}{p^{2}}p^{2}d\Omega
=-\Omega \,,  \label{solidangle}
\end{eqnarray}%
where $\Omega $ is the solid angle of the integrated area. The result should
be valid as long as the integrated area and its boundary avoid singularities
and branch cuts. Since it is the curl of ${\bf a}$ that appears in the
surface integration (\ref{solidangle}), gauge transformations ${\bf %
a\rightarrow a}+\nabla \chi _{{\bf p}}$ will not change the result if $\chi
_{{\bf p}}$ is a single-valued function of ${\bf p.}$ This is consistent
with the line integral (\ref{Berrysphase}) since for any single-valued $\chi
_{{\bf p}},$ the difference $\Delta \chi _{{\bf p}}$ vanishes over a closed
loop.

If, as in the optical-fiber experiments \cite{Chiao86}, ${\bf p}$ describes
a circular path that makes a fixed angle $\theta $ with any fixed direction,
then $\Omega =2\pi \left( 1-\cos \theta \right) $ so that the change in
phase angle is 
\begin{equation}
\gamma _{\kappa }=\kappa \oint d{\bf p}\cdot {\bf a}^{\left( 1\right) }=2\pi
\kappa \left( \cos \theta -1\right) \text{.}  \label{DeltachiB}
\end{equation}%
However, the derivation also apparently works if ${\bf a}^{\left( 1\right) }$
is replaced by ${\bf a}^{\left( 0\right) },$ even though these results
should differ by $\kappa \Delta \chi _{{\bf p}}^{\left( 1\right) }=-2\pi
\kappa $ for a closed loop around ${\bf e}_{3}\,.$ While such a difference
is not observable, we can trace its origin to the nonintegrable nature of $%
\chi _{{\bf p}}^{\left( 1\right) }$ and the associated string singularity
through the integrated area of the surface integral (\ref{solidangle}). The
argument can in fact be turned around to imply that because the difference
cannot be observable, the helicity must be quantized.

Since the string contribution is frequently ignored, it may be useful to
elaborate its role. Recall that Dirac's magnetic monopole contains a string
to bring the magnetic flux to the monopole inside a solenoid with a
vanishing diameter. Although strings are not evident in our expression (\ref%
{curla}) of $\nabla \times {\bf a,}$ their presence is implied by the
expression (\ref{a}) for ${\bf a}^{\left( 0\right) }.$ The physics is
clearer if we write ${\bf a}^{\left( 0\right) }$ as the limit of a
nonsingular function:%
\begin{equation}
{\bf a}^{\left( 0\right) }=\bbox{\hat{\phi}}\lim_{p_{0}\rightarrow 0}\frac{%
p_{\bot }\cos \theta }{p_{\bot }^{2}+p_{0}^{2}}\,,
\end{equation}%
where $p_{\bot }\equiv p\sin \theta $ is the momentum-space distance from
the ${\bf e}_{3}$ axis and $p_{0}$ represents the approximate diameter of
the solenoid. The ``magnetic field'' corresponding to the ``vector
potential'' is%
\begin{equation}
\nabla \times {\bf a}^{\left( 0\right) }=\lim_{p_{0}\rightarrow 0}\left[ -%
{\bf \hat{p}}\frac{\sin ^{2}\theta }{p_{\bot }^{2}+p_{0}^{2}}+{\bf e}_{3}%
\frac{2p_{0}^{2}\cos \theta }{\left( p_{\bot }^{2}+p_{0}^{2}\right) ^{2}}%
\right] \,.
\end{equation}%
The first term becomes the monopole field in the limit, and the second term
represents the two axial strings. It vanishes everywhere except on the axis
and can be expressed as a two-dimensional Dirac delta function, giving%
\begin{equation}
\nabla \times {\bf a}^{\left( 0\right) }=-\frac{{\bf \hat{p}}}{p^{2}}+2\pi 
{\bf e}_{3}\cos \theta \delta ^{\left( 2\right) }\left( p_{\bot }\right) \,,
\end{equation}%
where we can also write $\delta ^{\left( 2\right) }\left( p_{\bot }\right)
=\delta \left( p_{1}\right) \delta \left( p_{2}\right) ,$ and the
coefficient of the second term on the rhs has been chosen to give the
correct surface integral over $p_{\bot }dp_{\bot }d\phi $ at fixed $\left|
p\cos \theta \right| \gg p_{0}.$ On the string, $\cos \theta =\pm 1,$ so
that we have two half strings along the two halves of the ${\bf e}_{3}$
axis, both taking ``magnetic flux'' away from the origin. The string term
adds exactly $2\pi $ to the surface integral (\ref{solidangle}), thereby
restoring Stokes theorem and bringing the surface and line integrals into
agreement. As mentioned above, gauge transformations with various functional
forms $\chi _{{\bf p}}\left( \theta ,\phi \right) $ can shift the strings
and replace them by explicitly nonintegrable functions. The choice of $\chi
_{{\bf p}}^{\left( 1\right) }$ makes ${\bf a}$ nonsingular over the
integrated surface in Eq.\thinspace (\ref{solidangle}) between the loop and
the upper pole $\theta =0$, but ${\bf a}^{\left( 0\right) }$ has a
singularity penetrating the same surface, giving both line and surface
integrals for the phase difference $-\kappa \left( \Delta \chi -\Delta \chi
_{{\bf p}}\right) $ that differ by $\kappa {\bf \Delta }\chi _{{\bf p}%
}^{\left( 1\right) }.$ The agreement also extends to use of the surface that
includes the pole at $\theta =\pi $\thinspace for both cases 0 and 1. For
the choice $\chi _{{\bf p}}^{\left( 2\right) },$ the line and surface
integrals both vanish if the branch cut is inserted explicitly in the
surface integral, so that the nonintegrable function in Eq.\thinspace (\ref%
{a2}) is replaced by%
\begin{equation}
{\bf a}^{\left( 2\right) }=\frac{\bbox{\hat{\theta }}}{p}\left[ \phi -2\pi
h\left( \phi -\phi _{0}\right) \right] \sin \theta \,,
\end{equation}%
for example, with $0<\phi _{0}<2\pi $ where $h\left( \phi -\phi _{0}\right) $
is the Heaviside step function.

\section{Discussion}

In this Section, the nonexistence proofs and recent papers concerning
localized states are briefly reviewed, the new photon position operators are
discussed in the context of this literature, and our results are summarized.

The most quoted paper is that of Newton and Wigner \cite{NewtonWigner}.
These authors assumed a rotationally invariant set of localized states and
arrived at the position operator of the form $ip^{\alpha }\nabla p^{-\alpha
} $ for spinless particles with or without mass. They also obtained an
expression of the position operator for massless particles of spin $\frac{1}{%
2}.$ Regarding photons, they stated that for $S=1$ and higher ``we found
that no localized states in the above sense exist. This is an
unsatisfactory, if not unexpected, feature of our work.'' As a result of
their conclusion, it is frequently stated that a (spatial) photon wave
function does not exist \cite{CohenTannoudji}. The localization postulate
adopted by Newton and Wigner is strong (any displacement of a localized
state is assumed to make it orthogonal to states of the undisplaced set) and
has been the focus of a number of more recent studies. In particular,
Wightman and others \cite{Wightman} have used generalized imprimitivities %
\cite{Mackey} to reformulate localization more rigorously in terms of
localizability in a region. However, their work did not alter the conclusion
that a single photon is not localizable. Other authors have sought effective
wave functions that satisfy a somewhat relaxed localization condition \cite%
{Sipe95,Mourad,BB2,BB3,APS}.

We suggest a different potential problem with the conclusion of Newton and
Wigner for massless particles of spin $S\geq 1$. To ensure a rotationally
invariant linear manifold of localized states for a system with total
angular momentum quantum number $j\,$, they assumed a {\em complete set} of $%
2j+1$ wave functions $\psi _{jm},\,-j\leq m\leq j\,,$ where $m$ is a
component referenced to an external direction. While the existence of a
complete set is sufficient to give a rotationally invariant manifold, it is
not necessary for massless particles of spin $S>\frac{1}{2}.$ Massless
particles with spin have only two spin states, namely those corresponding to
the helicities $\pm S.$ For a system of states at the coordinate origin, the
orbital angular momentum vanishes and $j=S.$ The states in the linear
manifold are characterized by components of ${\bf j}$ not along a
space-fixed direction but along the momentum direction ${\bf \hat{p}\,.}$
For $S>\frac{1}{2},$ the manifold is not complete and consequently it cannot
describe a state with spin quantized along an {\em arbitrary direction}.
However, it {\em can} describe the allowed states with either helicity.
Furthermore, since the helicity operator commutes with the generator ${\bf J}
$ of rotations, the two helicity subspaces {\em are separately rotationally
invariant}. Because the helicity eigenstates form a complete rotational set
only for $S\leq \frac{1}{2},$ it is clear why the Newton's and Wigner's
insistence on a complete rotational manifold is stronger than necessary for
massless particles with $S\geq 1\,.$

In the papers based on the method of generalized imprimitivities \cite%
{Wightman}, it appears that the system of commuting imprimitivities, through
which the position operator is defined, is assumed to be independent of
spin. (This arises from Wightman's axiom V \cite{Wightman}). As seen above,
for example in Eq.\thinspace (\ref{rrP}), the spin is inextricably coupled
to the momentum and thereby to the position operator. Position operators are
known to exist for massless particles of spin 0, and it may be possible to
find a system of imprimitivities like that for the scalar case for each
value of the invariant helicity. In any case, the imprimitivities must be
distinct for different helicities because the position operators are.

In several more recent proofs of the nonexistence of a photon position
operator with commuting components, operator algebra was used with the
assumption that the position operators satisfy Eq.\/(\ref{commJ}) \cite%
{JordanMukunda,Jadczyk73,Jordan,Mourad}. However, we have shown above and in
Appendix C that the matrix \underline{${\bf r}$} includes a ``gauge
potential'' ${\bf a}$ that is transformed by rotations. As a result, the
commutator $\left[ J_{j},r_{k}\right] $ contains an additional term
involving the spin, and indeed that this extra term is required in order to
give the correct phase of the rotated photon state. The components of the
position operators \/(\ref{rralph}) thus satisfy Eq.\/(\ref{rJcomm}), which
unlike the more familiar relation (\ref{commJ}), is compatible with
commuting components. Thus, nonexistence proofs that assume Eq.\thinspace (%
\ref{commJ}) do not apply.

Our construction of the photon position operator uses a spin rotation to
decouple the spin from the momentum while the gradient operator acts to
generate momentum-space translations. It seems to be the most natural way to
retain the usual role of the position operator in momentum space. While we
have not touched on important questions about the limits to which the photon
position is observable in a dynamic measurement, our demonstration that a
photon position operator does in fact exist means that there is no
nonvanishing commutator $\left[ \underline{r}_{j},\underline{r}_{k}\right] $
to limit the calculation of photon probability amplitudes. Simultaneous
eigenvectors of the operators $\underline{r}_{j}$ are available for
calculation of the probability that the corresponding eigenvalue is
observed. Of course limitations arising from Fourier analysis, similar to
those applicable to massive particles, still apply. Our work therefore
supports the view that photons have wave functions that are not
qualitatively different from those of massive particles, as concluded by Bia%
\l ynicki-Birula \cite{BB3} and Sipe \cite{Sipe95}. The rules of quantum
mechanics require that each observable be represented by a Hermitian
operator. The localized basis sets found here makes it possible to treat
photons like massive particles in quantum calculations of interference
experiments and other situations where particle amplitudes in coordinate
space are useful; in both cases the usual rules of quantum mechanics can be
applied.

There is no unique representation of the photon position operator. Just as
there are many spin bases that can be used to describe the internal state of
a massive particle, there are many bases that can be used to describe the
combined translational and internal motion of a photon. The peculiar aspect
of the photon is that, because of the coupling of the spin to the momentum,
the position operator is a matrix that does not commute with the spin.
Different position operators with commuting components can be defined by
specifying the dependence $\chi _{{\bf p}}\left( \theta ,\phi \right) $ of
the axial rotation angle on the polar angles of ${\bf \hat{p}.}$ Different
choices of $\chi _{{\bf p}}$ lead to different ``gauge potentials'' ${\bf a}$
for the phase of the photon state at different positions in momentum space.
In general, phase changes depend on the path and are thus described by a
nonintegrable function. While a specific choice of gauge usually results in
operator expressions with less than the expected symmetry of the
Hamiltonian, the symmetry is obtained within the full group of possible
gauges. The forms that result from a selection of $\chi _{{\bf p}}$ are all
unitarily equivalent to each other and to the position operator for a
massive particle, $ip^{\alpha }\nabla p^{-\alpha }$, where the unitary
operator is the rotation through Euler angles. The operator $ip^{\alpha
}\nabla p^{-\alpha }$ has eigenvectors with a fixed direction in space and
is independent of the spin. Consequently, for massive particles the
description of spin is a separable problem. However, the spin and momentum
of a photon are inexorably coupled, since the direction of ${\bf p}$
determines the direction of the observable component ${\bf \hat{p}\cdot }$%
\underline{${\bf S}$} of the internal angular momentum. Position operators
with transverse and longitudinal eigenvectors could in principle be used to
describe a massive particle, but this is probably not useful since $%
ip^{\alpha }\nabla p^{-\alpha }$ is a simpler alternative. For a massless
particle this choice does not exist, reflecting the fact that for a photon,
the orbital and spin angular momenta are not separable.

In summary, the arguments presented here show that a photon is much like any
other particle in that its position is an observable described by a set of
three commuting Hermitian operators. However, the photon (as well as other
massless particles of spin $S>\frac{1}{2}\,$) has only two linearly
independent spin states, and in these states the spin is coupled to the
momentum. As a result, its position operator is a matrix that does not
commute with the spin. Different selections of the function $\chi _{{\bf p}%
}\left( \theta ,\phi \right) $ generally give different position operators,
so that the position operator is not unique and does not transform under $%
{\bf J}$ as a simple vector. However, the eigenvectors of any one of these
unitarily equivalent position operators gives a basis of localized states
with unique eigenvalues that are independent of helicity, and there is
consequently no disagreement as to the actual position of the photon.
Contrary to the traditional view, localized basis sets do exist, and it
appears that photon wave functions can be defined according the usual rules
of quantum mechanics.

\section*{Acknowledgments}

The authors wish to thank the Natural Sciences and Engineering Research
Council of Canada for financial support. It is also a pleasure to
acknowledge stimulating and helpful communication from B.-S. Skagerstam. One
of us (W. E. B.) thanks A. Lasenby and the Astrophysics Group of the
Cavendish Laboratory, University of Cambridge, for hospitality during a
sabbatical leave taken there, and he thanks J. P. Crawford for helpful
discussions of string singularities.\appendix

\section{Decomposition of Rotations}

The purpose of this appendix is to derive relations between an infinitesimal
rotation about an arbitrary axis in three-dimensional space and angular
parameters in a product of rotations around specified axes. These relations
are then used to predict rotational properties of the photon position
operator and its eigenstates. The angular parameters are the Euler angles
that specify the polar and azimuthal coordinates $\theta $ and $\phi ,$
respectively, of a given direction ${\bf \hat{p}}$ together with an axial
angle $\chi $ about ${\bf \hat{p}}$. One can express the Euler-angle
parametrization in terms of active rotations about spaced-fixed axes: with
the ${\bf p}$-frame axes initially coincident with the space-fixed axes $%
{\bf e}_{1},{\bf e}_{2},{\bf e}_{3},$ of the lab frame, a first rotation by $%
\chi $ about ${\bf e}_{3}$ is followed by a rotation by $\theta $ about $%
{\bf e}_{2}$ and finally a rotation by $\phi $ about ${\bf e}_{3}.$ The
axial rotation angle $\chi $ does not affect the direction ${\bf \hat{p}}$
and may be chosen to be a function, say $\chi _{{\bf p}}\left( \theta ,\phi
\right) ,$ of the ``local coordinates'' $\theta ,\phi $ that specify ${\bf 
\hat{p}}${\bf . }A rotation by the Euler angles $\left( \phi ,\theta ,\chi _{%
{\bf p}}\left( \theta ,\phi \right) \right) $ has only two degrees of
freedom and is uniquely determined by the direction ${\bf \hat{p}.}$

The derivation is easily constructed using tools of Clifford's geometric
algebra $C\!\ell _{3}$ of 3-dimensional space \cite%
{Hestenes66,Baylis99,Lounesto97}. Those not familiar with the algebra can
follow the derivation in terms of the $2\times 2$ matrices that form the
common matrix representation in which the Cartesian unit vectors ${\bf e}%
_{1},{\bf e}_{2},{\bf e}_{3},$ are replaced by the corresponding Pauli spin
matrices familiar to physicists. In the algebra, vectors are rotated by
transformations of the form%
\begin{equation}
{\bf v}\rightarrow R{\bf v}R^{\dag },
\end{equation}%
where $R\left( \bbox{\xi }\right) =\exp \left( -i\bbox{\xi
}/2\right) \in SU\left( 2\right) $ is the element for a rotation by the
angle $\xi =\left| \bbox{\xi }\right| $ about the axis ${\bf \hat{\xi}}$ .
The rotation specified by the Euler angles $\left( \phi ,\theta ,\chi
\right) $ is given by the rotation element%
\begin{align}
{\cal R}\left( \phi ,\theta ,\chi \right) & \equiv R\left( \phi {\bf e}%
_{3}\right) R\left( \theta {\bf e}_{2}\right) R\left( \chi {\bf e}_{3}\right)
\\
& =\exp \left( -i\phi {\bf e}_{3}/2\right) \exp \left( -i\theta {\bf e}%
_{2}/2\right) \exp \left( -i\chi {\bf e}_{3}/2\right) .  \nonumber
\end{align}%
An additional rotation will generally change all three angular parameters $%
\phi ,\theta ,\chi \,.$ We want to determine the changes caused by an
infinitesimal rotation $R\left( d\bbox{\xi }\right) {\bf \,.}$

To clarify our objective, we first consider the simple case in which $d%
\bbox{\xi }{\bf =}d\xi \,{\bf e}_{3}\,.$ The only effect of $R\left( d%
\bbox{\xi }\right) $ is to increment the azimuthal angle $\phi $ :%
\begin{equation}
{\cal R}\left( \phi ,\theta ,\chi \right) \rightarrow R\left( d\xi {\bf e}%
_{3}\right) {\cal R}\left( \phi ,\theta ,\chi \right) ={\cal R}\left( \phi
+d\xi ,\theta ,\chi \right) \,.
\end{equation}%
This result is easily written in terms of the rotations ${\cal R}\left( \phi
,\theta ,\chi _{{\bf p}}\left( \theta ,\phi \right) \right) $ that have only
two degrees of freedom:%
\begin{eqnarray}
&&R\left( d\xi {\bf e}_{3}\right) {\cal R}\left( \phi ,\theta ,\chi _{{\bf p}%
}\left( \theta ,\phi \right) \right)  \label{dxiR} \\
&=&{\cal R}\left( \phi +d\xi ,\theta ,\chi _{{\bf p}}\left( \theta ,\phi
+d\xi \right) \right) R\left( -{\bf e}_{3}d\chi _{{\bf p}}\right) \,, 
\nonumber
\end{eqnarray}%
where $\chi _{{\bf p}}\left( \theta ,\phi +d\xi \right) =\chi _{{\bf p}%
}\left( \theta ,\phi \right) +d\chi _{{\bf p}}$ and $d\chi _{{\bf p}}=d\xi
\,\partial \chi _{{\bf p}}/\partial \phi \,.$

Now we generalize this approach to an arbitrary infinitesimal rotation by $d%
\bbox{\xi }=d\xi _{1}{\bf e}_{1}+d\xi _{2}{\bf e}_{2}+d\xi _{3}{\bf e}%
_{3}\,. $We initially consider $\chi $ an independent parameter and solve $%
R\left( d\bbox{\xi }\right) {\cal R}\left( \phi ,\theta ,\chi \right) ={\cal %
R}\left( \phi +d\phi ,\theta +d\theta ,\chi +d\chi \right) $ for the
infinitesimal changes $d\phi ,d\theta ,d\chi $ in the Euler angles. To first
order in the changes, ${\cal R}\left( \phi +d\phi ,\theta +d\theta ,\chi
+d\chi \right) $%
\begin{align}
& =\left[ 1-\frac{i}{2}\left( {\bf e}_{3}d\phi +e^{-i{\bf e}_{3}\phi }{\bf e}%
_{2}d\theta +R{\bf e}_{3}R^{\dag }d\chi \right) \right] {\cal R}\left( \phi
,\theta ,\chi \right)  \nonumber \\
& =\left\{ 1-\frac{i}{2}\left[ {\bf e}_{1}\left( \sin \theta \cos \phi d\chi
-\sin \phi d\theta \right) +{\bf e}_{2}\left( \cos \phi d\theta \right.
\right. \right.  \label{dchiR} \\
& \left. \vspace{0.08in}\left. \left. +\sin \theta \sin \phi d\chi \right) +%
{\bf e}_{3}\left( d\phi +\cos \theta d\chi \right) \right] \right\} {\cal R}%
\left( \phi ,\theta ,\chi \right) \,,  \nonumber
\end{align}%
in which we have explicitly accounted for the lack of commutivity of
rotations about different axes. Equation (\ref{dxiR}) therefore implies%
\begin{eqnarray}
d\bbox{\xi } &=&{\bf e}_{1}\left( \sin \theta \cos \phi d\chi -\sin \phi
d\theta \right) \\
&&+{\bf e}_{2}\left( \cos \phi d\theta +\sin \theta \sin \phi d\chi \right) +%
{\bf e}_{3}\left( d\phi +\cos \theta d\chi \right) \,,  \nonumber
\end{eqnarray}%
which is readily solved to give%
\begin{align}
d\phi & =d\xi _{3}-\cot \theta \left( d\xi _{1}\cos \phi +d\xi _{2}\sin \phi
\right)  \nonumber \\
d\theta & =d\xi _{2}\cos \phi -d\xi _{1}\sin \phi  \label{dphithetachi1} \\
d\chi & =\frac{d\xi _{1}\cos \phi +d\xi _{2}\sin \phi }{\sin \theta }. 
\nonumber
\end{align}%
Note the singular nature of the relations for $d\phi $ and $d\chi $ in the
limit $\theta \rightarrow 0\,.$ Singularities are common whenever general
rotations are parametrized in terms of rotations about specified axes. They
are related to the nonuniqueness of the parametrization for some rotations,
for example all of the rotations $D\left( \phi ,0,\chi -\phi \right) $ for
fixed $\chi $ and arbitrary $\phi $ are equal. We can also express the
rotation (\ref{dchiR}) in terms of rotations with two degrees of freedom%
\begin{align}
& R\left( d\bbox{\xi }\right) {\cal R}\left( \phi ,\theta ,\chi _{{\bf p}%
}\right) ={\cal R}\left( \phi +d\phi ,\theta +d\theta ,\chi _{{\bf p}}+d\chi
\right)  \nonumber \\
& ={\cal R}\left( \phi +d\phi ,\theta +d\theta ,\chi _{{\bf p}}+d\chi _{{\bf %
p}}\right) R\left( \left( d\chi -d\chi _{{\bf p}}\right) {\bf e}_{3}\right)
\label{RRRR} \\
& =R\left( \left( d\chi -d\chi _{{\bf p}}\right) {\bf \hat{p}}\right) {\cal R%
}\left( \phi +d\phi ,\theta +d\theta ,\chi _{{\bf p}}+d\chi _{{\bf p}}\right)
\nonumber
\end{align}%
where $\chi _{{\bf p}}=\chi _{{\bf p}}\left( \theta ,\phi \right) $ and%
\begin{align}
d\chi -d\chi _{{\bf p}}& =d\chi -\left( \frac{\partial \chi _{{\bf p}}}{%
\partial \phi }d\phi +\frac{\partial \chi _{{\bf p}}}{\partial \theta }%
d\theta \right) \\
& =\left[ \frac{\cos \phi }{\sin \theta }\left( 1+\cos \theta \frac{\partial
\chi _{{\bf p}}}{\partial \phi }\right) +\sin \phi \frac{\partial \chi _{%
{\bf p}}}{\partial \theta }\right] d\xi _{1}  \nonumber \\
& +\left[ \frac{\sin \phi }{\sin \theta }\left( 1+\cos \theta \frac{\partial
\chi _{{\bf p}}}{\partial \phi }\right) -\cos \phi \frac{\partial \chi _{%
{\bf p}}}{\partial \theta }\right] d\xi _{2}  \nonumber \\
& -\frac{\partial \chi _{{\bf p}}}{\partial \phi }d\xi _{3}
\label{dchidchip} \\
& =\left( {\bf a\times p+\hat{p}}\right) \cdot d\bbox{\xi }\,,  \label{pap}
\end{align}%
where in the last step we used the definition (\ref{a},\ref{a0}) of ${\bf %
a\,.}$ While the derivation employs tools of geometric algebra, the result
is generally valid for spatial rotations in three-dimensional space.

A similar analysis gives the result of an infinitesimal rotations added to
the rhs:%
\begin{align}
& {\cal R}\left( \phi ,\theta ,\chi _{{\bf p}}\right) R\left( d\bbox{\xi }%
^{\prime }\right) ={\cal R}\left( \phi +d\phi ,\theta +d\theta ,\chi _{{\bf p%
}}+d\chi \right) \\
& ={\cal R}\left( \phi +d\phi ,\theta +d\theta ,\chi _{{\bf p}}+d\chi _{{\bf %
p}}\right) R\left( \left( d\chi -d\chi _{{\bf p}}\right) {\bf e}_{3}\right)
\,.  \label{RRRR2}
\end{align}%
In this case, we obtain%
\begin{align}
d\phi & =\frac{-d\xi _{1}^{\prime }\cos \chi +d\xi _{2}^{\prime }\sin \chi }{%
\sin \theta }  \nonumber \\
d\theta & =d\xi _{1}^{\prime }\sin \chi +d\xi _{2}^{\prime }\cos \chi
\label{dphithetachi} \\
d\chi & =d\xi _{3}^{\prime }-d\phi \cos \theta  \nonumber \\
d\chi -d\chi _{{\bf p}}& =d\xi _{1}^{\prime }\left[ \frac{\cos \chi }{\sin
\theta }\left( \cos \theta +\frac{\partial \chi _{{\bf p}}}{\partial \phi }%
\right) -\sin \chi \frac{\partial \chi _{{\bf p}}}{\partial \theta }\right] 
\nonumber \\
& -d\xi _{2}^{\prime }\left[ \frac{\sin \chi }{\sin \theta }\left( \cos
\theta +\frac{\partial \chi _{{\bf p}}}{\partial \phi }\right) +\cos \chi 
\frac{\partial \chi _{{\bf p}}}{\partial \theta }\right]  \nonumber \\
& +d\xi _{3}^{\prime }\,.
\end{align}%
These results can also be obtained from the Hermitian conjugate of the
relations (\ref{dchiR}) and (\ref{dxiR}) by noting that $D^{\dag }\left(
\phi ,\theta ,\chi \right) =D\left( -\chi ,-\theta ,-\phi \right) $. Such
added rotations from the rhs are rotations in the ${\bf p}$-frame and can
also be expressed by%
\begin{equation}
{\cal R}\left( \phi ,\theta ,\chi _{{\bf p}}\right) R\left( d\xi
_{j}^{\prime }{\bf e}_{j}\right) =R\left( d\xi _{j}^{\prime }{\bf e}_{{\bf p}%
j}\right) {\cal R}\left( \phi ,\theta ,\chi _{{\bf p}}\right) \,.
\label{rhstolhs}
\end{equation}%
Evidently the Cartesian components of $d\bbox{\xi }^{\prime }$ are just the $%
{\bf p}$-frame components of $d\bbox{\xi }$, that is, $d\xi _{j}^{\prime
}=d\xi _{{\bf p}j}\,.$ We make use of this result in Appendix D$.$

\section{Application to rotations of state vectors}

The helicity state vectors \underline{${\bf e}_{{\bf p}\kappa }$} are
obtained by rotating the constant column vectors \underline{${\bf e}_{\kappa
}$}%
\begin{equation}
\underline{{\bf e}_{{\bf p}\kappa }}=\underline{D}\left( \phi ,\theta ,\chi
\right) \underline{{\bf e}_{\kappa }}\,,  \label{epk}
\end{equation}%
by the rotation matrix%
\begin{equation}
\underline{D}=\exp \left( -i\underline{S_{3}}\phi \right) \exp \left( -i%
\underline{S_{2}}\theta \right) \exp \left( -i\underline{S_{3}}\chi \right)
\end{equation}%
generated by the spin-one matrices \underline{$S_{j}$}. Starting in the
Cartesian basis with%
\begin{equation}
\underline{{\bf e}_{0}}=\left( 
\begin{array}{c}
0 \\ 
0 \\ 
1%
\end{array}%
\right) ,\;\underline{{\bf e}_{\pm 1}}=\frac{1}{\sqrt{2}}\left( 
\begin{array}{c}
1 \\ 
\pm i \\ 
0%
\end{array}%
\right) ,
\end{equation}%
the rotation matrix is explicitly\widetext%
\begin{equation}
D=\left( 
\begin{array}{ccc}
\cos \theta \cos \phi \cos \chi -\sin \phi \sin \chi & -\left( \sin \phi
\cos \chi +\cos \theta \cos \phi \sin \chi \right) & \sin \theta \cos \phi
\\ 
\cos \theta \sin \phi \cos \chi +\cos \phi \sin \chi & \cos \phi \cos \chi
-\cos \theta \sin \phi \sin \chi & \sin \theta \sin \phi \\ 
-\sin \theta \cos \chi & \sin \theta \sin \chi & \cos \theta \,%
\end{array}%
\right) ,  \label{Dmatrix}
\end{equation}%
and we thus find\narrowtext%
\begin{eqnarray}
\underline{{\bf e}_{{\bf p}0}} &=&\left( 
\begin{array}{c}
\sin \theta \,\cos \phi \\ 
\sin \theta \,\sin \phi \\ 
\cos \theta \,%
\end{array}%
\right) \,, \\
\underline{{\bf e}_{{\bf p}\kappa }} &=&\frac{e^{-i\kappa \chi }}{\sqrt{2}}%
\left( 
\begin{array}{c}
\cos \theta \cos \phi -i\kappa \sin \phi \\ 
\cos \theta \sin \phi +i\kappa \cos \phi \\ 
-\sin \theta%
\end{array}%
\right) \,,\;\kappa =\pm 1\,.
\end{eqnarray}%
Note that $\underline{{\bf e}_{{\bf p}0}}$ is a matrix representation of $%
{\bf \hat{p},}$ and $\underline{{\bf e}_{{\bf p}\kappa }}$ depends on $\chi $
simply through the phase factor $e^{-i\kappa \chi }.$ We put $\chi =\chi _{%
{\bf p}}\left( \theta ,\phi \right) $ to obtain vectors \underline{${\bf e}_{%
{\bf p}\kappa }$} that depend only on the two angles $\theta $ and $\phi ,$
and hence only on the direction of ${\bf \hat{p}\,.}$ The direction of ${\bf 
\hat{p}}$ can be changed by a further rotation of \underline{${\bf e}_{{\bf p%
}\kappa }$} by $d\bbox{\xi }$ of the form $\exp \left( -id\bbox{\xi \cdot }%
\underline{{\bf S}}\right) \underline{{\bf e}_{{\bf p}\kappa }}$%
\begin{align}
& =\underline{D}\left( \phi +d\phi ,\theta +d\theta ,\chi _{{\bf p}}+d\chi
\right) \underline{{\bf e}_{\kappa }} \\
& =\exp \left[ -i\kappa \left( d\chi -d\chi _{{\bf p}}\right) \right] 
\underline{D}\left( \phi +d\phi ,\theta +d\theta ,\chi _{{\bf p}}+d\chi _{%
{\bf p}}\right) \underline{{\bf e}_{\kappa }}  \nonumber \\
& =\exp \left[ -i\kappa \left( d\chi -d\chi _{{\bf p}}\right) \right] 
\underline{{\bf e}_{{\bf p}+d{\bf p}\kappa }}\,.
\end{align}%
The rotation matrix $\exp \left( -id\bbox{\xi \cdot }\underline{{\bf S}}%
\right) $ shuffles the components of the vector $\underline{{\bf e}_{{\bf p}%
\kappa }}.$ A complete rotation of $\underline{{\bf e}_{{\bf p}\kappa }}$
also includes the operator $\exp \left( -id\bbox{\xi
\cdot L}\right) $ which changes the angular arguments $\theta ,\phi $ so as
to rotate ${\bf \hat{p}}$ to ${\bf \hat{p}}-d{\bf \hat{p}}$ . We thus find $%
\exp \left( -id\bbox{\xi
\cdot }\underline{{\bf J}}\right) \underline{{\bf e}_{{\bf p}\kappa }}$%
\begin{align}
& =\exp \left[ -i\kappa \left( d\chi -d\chi _{{\bf p}}\right) \right] \exp
\left( -id\bbox{\xi \cdot L}\right) \underline{{\bf e}_{{\bf p}+d{\bf p}%
\kappa }}\quad \quad \\
& =\exp \left[ -i\kappa \left( d\chi -d\chi _{{\bf p}}\right) \right] 
\underline{{\bf e}_{{\bf p}\kappa }}  \label{phasechange}
\end{align}%
and consequently the only effect of a rotation of $\underline{{\bf e}_{{\bf p%
}\kappa }}$ by $d\xi $ is to change its phase by $-\kappa \left( d\chi
-d\chi _{{\bf p}}\right) \,,$ where $d\chi -d\chi _{{\bf p}}$ is given by
Eq. (\ref{dchidchip}). This agrees with the partition of \underline{${\bf J}$%
} into%
\begin{equation}
\underline{{\bf J}}=\underline{{\bf L}}^{\left( r\right) }+\underline{{\bf S}%
}^{\left( r\right) }
\end{equation}%
since 
\begin{equation}
\underline{{\bf L}}^{\left( r\right) }\underline{{\bf e}_{{\bf p}\kappa }}=%
\underline{D}\,{\bf L}\underline{D}^{-1}\underline{{\bf e}_{{\bf p}\kappa }}=%
\underline{D}\,{\bf L}\underline{{\bf e}_{\kappa }}=0
\end{equation}%
and%
\begin{align}
\underline{{\bf S}^{\left( r\right) }}& ={\bf \hat{p}\cdot }\underline{{\bf S%
}}\left( {\bf a\times p+\hat{p}}\right) \quad \quad \\
{\bf \hat{p}\cdot }\underline{{\bf S}}\,\underline{{\bf e}_{{\bf p}\kappa }}%
& =\underline{D}\underline{S_{3}}\underline{D}^{-1}\underline{{\bf e}_{{\bf p%
}\kappa }}=\underline{D}\underline{S_{3}}\,\underline{{\bf e}_{\kappa }}%
=\kappa \underline{{\bf e}_{{\bf p}\kappa }} \\
\left( {\bf a\times p+\hat{p}}\right) \cdot d\bbox{\xi }& =d\chi -d\chi _{%
{\bf p}}\,.
\end{align}

\section{Application to rotations of the position operator}

The new position operator has components%
\begin{equation}
\underline{r_{l}}=ip^{\alpha }\underline{D}\frac{\partial }{\partial p_{l}}%
\underline{D}^{-1}p^{-\alpha }.
\end{equation}%
The components of vector operators are expected to rotate into one another
as given by the commutation relation (\ref{commJ}). However, an additional
factor arises by the need to transform the matrix associated with each
vector component by an axial rotation. The required spin rotation is given
by that for the rotation matrix \underline{$D$}$.$ From relation (\ref{RRRR}%
) we see that the rotation transforms $\underline{D}\left( \phi ,\theta
,\chi _{{\bf p}}\right) $ to $\left( 1-i\underline{{\bf S}}\cdot d\bbox{\xi }%
\right) \underline{D}\left( \phi ,\theta ,\chi _{{\bf p}}\right) $%
\begin{align}
& =\underline{D}\left( \phi +d\phi ,\theta +d\theta ,\chi _{{\bf p}}+d\chi _{%
{\bf p}}\right) \left( 1-i\underline{S_{3}}\left( d\chi -d\chi _{{\bf p}%
}\right) \right) \\
& =\left( 1-i{\bf \hat{p}\cdot }\underline{{\bf S}}\left( d\chi -d\chi _{%
{\bf p}}\right) \right) \underline{D}\left( \phi +d\phi ,\theta +d\theta
,\chi _{{\bf p}}+d\chi _{{\bf p}}\right) ,
\end{align}%
where from Eqs. (\ref{pap}) and (\ref{dchidchip}),%
\begin{equation}
{\bf \hat{p}\cdot }\underline{{\bf S}}\left( d\chi -d\chi _{{\bf p}}\right) =%
{\bf \hat{p}\cdot }\underline{{\bf S}}\,d\bbox{\xi \cdot }\left( {\bf \hat{p}%
}+{\bf a\times p}\right) =\underline{{\bf S}}^{\left( r\right) }\cdot d%
\bbox{\xi }\,.
\end{equation}%
The result is to add $\left[ \underline{S_{j}}^{\left( r\right) },\underline{%
r_{k}}\right] $ to the rhs of Eq. (\ref{commJ}). The added term is required
for consistency with the phase changes (\ref{phasechange}) induced in the
basis vectors by rotations.

\section{Rotations of the reference frame and Berry's phase}

An extra rotation acting on the rhs of $R\left( \phi ,\theta ,\chi \right) $
is equivalent to the opposite rotation to the reference frame. As shown in
relation (\ref{rhstolhs}), it is also equivalent to a rotation from the left
by the same angles but about ${\bf p}$-frame axes instead of lab axes. This
relation is useful in deriving an expression for Berry's phase \cite{Berry}.
Berry's phase is a topological phase that arises from adiabatic transport of
a discrete state around a closed loop, and it is usually derived as a purely
quantum phenomenon \cite{Berry,Aharon,Anan}. For the polarized light,
however, it was shown by Haldane \cite{Haldane86} to be a consequence of
classical differential geometry. We derive it here by means of the classical
rotation operators introduced above.

Transport in the adiabatic limit corresponds to parallel transport in the
given gauge \cite{Chiao86,Simon}, which is equivalent to nonrotating
(Fermi-Walker) transport on geodesics. We achieve this in rotations of ${\bf %
p}$ by piecing together the closed loop from a sequence of great-circle
rotations, in each of which the rotation axis is orthogonal to ${\bf p}$ and
the axial rotation vanishes. Thus, in Eq.\thinspace (\ref{RRRR}) we vary $%
\xi ^{\prime }$ such that $d\xi _{3}^{\prime }=0\,.$ Then from Eq.\thinspace
(\ref{dphithetachi}),%
\begin{equation}
d\chi =-\cos \theta d\phi
\end{equation}%
We vary the parameters $\xi _{1}^{\prime }$ and $\xi _{2}^{\prime }$ so as
to complete a closed loop, increasing $\phi $ by $2\pi .$ The condition $%
d\xi _{3}^{\prime }=0$ ensures that the change in ${\bf \hat{p}}$ is always
along a great circle. There is a change in the axial angle $-\chi $ by 
\begin{equation}
-\Delta \chi =\oint \cos \theta \,d\phi  \label{chichange}
\end{equation}%
even after ${\bf p}$ has looped back to its original direction. If the
initial orientation of the photon frame is given by $R\left( \phi ,\theta
,\chi \right) $ with the momentum direction ${\bf \hat{p},}$ the orientation
after $\phi $ has increased by $2\pi $ is $R\left( \phi +2\pi ,\theta ,\chi
+\Delta \chi \right) ,$ giving a the net rotation of%
\begin{eqnarray}
&&R\left( \phi +2\pi ,\theta ,\chi +\Delta \chi \right) R^{-1}\left( \phi
,\theta ,\chi \right)  \nonumber \\
&=&\exp \left[ -i\left( \chi +\Delta \chi \right) {\bf \hat{p}}/2\right]
\exp \left( -i\pi {\bf e}_{3}\right) \exp \left( i\chi {\bf \hat{p}}/2\right)
\nonumber \\
&=&\exp \left[ -i\left( \Delta \chi +2\pi \right) {\bf \hat{p}}/2\right] ,
\end{eqnarray}%
where we used the symmetry%
\begin{equation}
R\left( \phi ,\theta ,\chi \right) =\exp \left( -i\chi {\bf \hat{p}}%
/2\right) R\left( \phi ,\theta ,0\right) ,
\end{equation}%
which allows us to cancel the $\theta $ rotations, and we noted that $\exp
\left( -i\pi {\bf e}_{3}\right) =-1=\exp \left( -i\pi {\bf \hat{p}}\right) .$
The total rotation under parallel transport about the loop is thus the
rotation about ${\bf \hat{p}}$ by the angle $\eta =\Delta \chi +2\pi =\oint
\left( 1-\cos \theta \right) d\phi .$

Note that the result does depend on the parallel transport of ${\bf p.}$ If
we rotate ${\bf p}$ directly around the space-fixed ${\bf e}_{3}$ axis, then
we simply change $\phi $ by $2\pi $ and the axial angle $\chi $ does not
change. The mathematical confirmation of this result is found from
Eq.\thinspace (\ref{dphithetachi1}) with $d\xi _{1}=d\xi _{2}=0.$ The
rotation in this case is not on a great circle (unless $\cos \theta =0$) and 
${\bf p}$ is therefore not parallel transported.

The phase change for electromagnetic plane waves of given helicity is
proportional to the rotation angle. To see the relation, consider
electromagnetic plane waves of helicity $\kappa ,$ for which the field ${\bf %
F}={\bf E}+ic{\bf B}$ is \cite{Baylis99}%
\begin{equation}
{\bf F}=\left( 1+{\bf \hat{p}}\right) {\bf E}_{0}\exp \left[ i\kappa \left(
\omega t-{\bf p\cdot x}\right) {\bf \hat{p}}\right] \,.
\end{equation}%
An additional axial rotation by $\eta $ transforms this to%
\begin{eqnarray}
&&\exp \left( -i\eta {\bf \hat{p}}/2\right) {\bf F}\exp \left( i\eta {\bf 
\hat{p}}/2\right) ={\bf F}\exp \left( i\eta {\bf \hat{p}}\right) \\
&=&\left( 1+{\bf \hat{p}}\right) {\bf E}_{0}\exp \left[ i\kappa \left(
\omega t-{\bf p\cdot x}+\kappa \eta \right) {\bf \hat{p}}\right]
\end{eqnarray}%
and gives the phase change $-\kappa \eta $ relative to $-\omega t.$ This is
exactly Berry's phase (\ref{Berrysphase})

\begin{equation}
\kappa \oint \left( \cos \theta -1\right) \,d\phi =-\kappa \Omega \,,
\end{equation}%
where $\Omega $ is the solid angle subtended by the loop. It is just that
calculated directly from the line integral in Section V when $\chi _{{\bf p}%
}=\chi _{{\bf p}}^{\left( 1\right) }=-\phi $ so that the rotation matrix $%
D\left( \phi ,\theta ,\chi _{{\bf p}}\right) $ rotates $S_{3}$ directly into 
${\bf S\cdot \hat{p},}$ and it agrees with the relation between angle and
geometric phase found by Berry \cite{Berry85}.

\end{document}